\documentclass[12pt, twoside]{article}
\usepackage{a4wide,amssymb,amsmath,amsfonts,bbm,cite}
\usepackage{epsfig,axodraw}

\newcommand{\be}{\begin{equation}}
\newcommand{\ee}{\end{equation}}
\newcommand{\bea}{\begin{eqnarray}}
\newcommand{\eea}{\end{eqnarray}}
\newcommand{\hsp}{\hspace{-2mm}/}
\newcommand{\cJ}{{\cal J}}
\newcommand{\bZ}{{\bf Z}}
\newcommand{\cO}{{\cal O}}

\begin{document}

\begin{titlepage}

\rightline{DESY-06-200}
\rightline{November 2006}

\begin{centering}
\vspace{1cm}
{\Large {\bf Gauge Coupling Unification in Six Dimensions }}\\

\vspace{1.5cm}

 {\bf Hyun Min Lee} \\
\vspace{.2in}

{\it Deutsches Elektronen-Synchrotron DESY, \\ Notkestra\ss e
        85, 22607 Hamburg, Germany} \\
\vskip 3mm        
{\it Department of Physics, Carnegie Mellon University, \\
5000 Forbes Avenue, Pittsburgh, PA 15213, USA}   


\end{centering}

\vspace{2cm}

\begin{abstract}
\noindent
We compute the one-loop gauge couplings in six-dimensional non-Abelian gauge theories
on the $T^2/Z_2$ orbifold with general GUT breaking boundary conditions. 
For concreteness, we apply the obtained general formulae to the gauge coupling running 
in a 6D $SO(10)$ orbifold GUT
where the GUT group is broken down to the standard model 
gauge group up to an extra $U(1)$. We find that the one-loop corrections
depend on the parity matrices encoding the orbifold boundary conditions 
as well as the volume and shape moduli of extra dimensions.
When the $U(1)$ is broken by the VEV of bulk singlets, the accompanying extra color
triplets also affect the unification of the gauge couplings. In this case, the $B-L$ breaking
scale compatible with the gauge coupling unification 
is sensitive to the change of the compactification scales.

\end{abstract}

\vspace{4.5cm}

\begin{flushleft}
E-mail address: hmlee@andrew.cmu.edu
\end{flushleft}

\end{titlepage}

%%%%%%%%%%%%%%%%%%%%%%%%%%%%%%%%%%%%%%%%%%%%%%%%%%%%%%%%%%%%%%%%%%%%
\section{Introduction}
%%%%%%%%%%%%%%%%%%%%%%%%%%%%%%%%%%%%%%%%%%%%%%%%%%%%%%%%%%%%%%%%%%%%
Grand Unified Theories(GUTs)\cite{guts} have been reconsidered recently in the context 
of orbifold GUTs\cite{oguts,kkl} where 
orbifold boundary conditions in extra dimensions are utilized 
to break down a GUT gauge symmetry to the Standard Model(SM) gauge
group and at the same time solve the doublet-triplet splitting problem.
In orbifold GUTs,
on top of the usual 4D logarithmic running to generate the difference between the SM
gauge couplings at low energy\cite{unification},
the Kaluza-Klein(KK) massive modes of the 4D gauge bosons 
give rise to additional threshold corrections. 
In orbifold GUTs, however, there is an ambiguity due to
the existence of the non-universal gauge kinetic terms localized
at the fixed points\cite{nibb,gls,lee} where the local gauge symmetry is reduced
compared to the bulk one.  
Nonetheless, by making a strong coupling assumption
at the cutoff scale\cite{naive}, the brane-localized gauge couplings may be ignored 
compared to the bulk gauge coupling, due to the volume
suppression of extra dimensions. So, the orbifold GUTs 
can provide a minimal setup for considering the threshold corrections 
consistent with a successful gauge coupling unification. 

Over the past years, 
the 5D orbifold GUTs have been much studied as a simplest case,
in particular, to compute the resulting effects 
of the KK massive modes to the gauge coupling unification 
for one flat\cite{dienes} or warped\cite{kchoi} extra dimension. 
For larger GUT groups such as $SO(10)$, however, the 5D case turns out not to be 
a minimal setup, 
because a usual Higgs mechanism is required for a further 
breaking to the SM gauge group
or unwanted massless modes of extra components of gauge fields must get massive\cite{so10}. 
In contrast, the 6D case has drawn more attention because it
is more economic to obtain the SM gauge group directly from a large GUT group\cite{abc,abc42} 
and there is more freedom to locate the SM fields for satisfying the experimental requirements
such as the flavor structure\cite{abcflavor} and the proton lifetime\cite{abcproton}. 
We may even regard the 6D orbifold GUTs as an effective theory of describing (heterotic) 
string compactifications as an intermediate GUT\cite{buchmstring}
with the hope to identify the remnants of string theory 
within the 6D orbifold field theory. 

In this paper, we consider the one-loop effective action for the gauge fields
containing zero modes in six-dimensional ${\cal N}=1$ supersymmetric 
GUTs compactified on a $T^2/Z_2$ orbifold. 
This can be regarded as a generalization of the previous findings on orbifolds
without gauge symmetry breaking\cite{gls}.
In the presence of the orbifold boundary conditions that are commuting, 
we obtain the bulk and brane contributions due to bulk vector and hyper multiplets 
and identify the necessary counterterms to cancel the divergences appearing 
in dimensional regularization. 
A bulk vector multiplet leads to both brane and bulk corrections 
while a bulk hyper multiplet gives rise only to a bulk 
correction. 
The bulk divergences are cancelled by a higher derivative term with universal coefficient 
whereas the brane divergences are cancelled by brane-localized gauge kinetic terms 
the coefficients of which depend on the local gauge symmetry at the fixed points.
In the case of the cutoff regularization\cite{gls}, there would be also power-like corrections 
in the cutoff scale to the gauge couplings,
but they don't affect the gauge coupling unification at all.
From the obtained effective action, we also derive the general expressions 
for the running of the effective gauge couplings for zero-mode gauge bosons.
In the low energy limit, we consider the running of the gauge couplings,
including the non-universal threshold corrections due to KK massive 
modes\footnote{For some early works on string theory computation of the one-loop gauge couplings,
see Ref.~\cite{stringhet,stringtype1}}.

We apply the general formulae for the gauge coupling running in
the six-dimensional $SO(10)$ orbifold GUT model proposed in Ref.~\cite{abc}.
This is the minimal setup to break $SO(10)$ down to the SM gauge group
up to a $U(1)$ factor
only by orbifold boundary conditions without obtaining massless modes from
the extra components of gauge bosons.
In some realistic $SO(10)$ orbifold GUT models,
we discuss about the possibility of having a large
volume of extra dimensions compatible
with the success of the gauge coupling unification.
We assume the breaking scale of the extra $U(1)$ to be lower than
the compactification scale in order to ignore the effect of the brane-localized
$U(1)$ breaking mass terms. 

For the case with isotropic compactification of extra dimensions, 
we show that the volume dependent
term of the KK threshold correction can give a sizable contribution to
the differential running of the gauge couplings for the large volume of extra dimensions. 
In this case, in order for the additional contribution
due to extra color triplets to be cancelled by the volume dependent part, 
the breaking scale of the extra $U(1)$ tends to be close to the compactification 
scale for the gauge coupling unification.
On the other hand, in the case with anisotropic compactification, e.g.  
in the 5D limit where the bulk gauge group becomes the Pati-Salam
$SU(4)\times SU(2)_L\times SU(2)_R$,
we show that the shape dependent term of the KK threshold correction
can be dominant, 
giving rise to the 5D power-like threshold corrections with non-universal coefficient 
in the compactification scales.
These power-like corrections in the 5D limit are {\it calculable},
in contrast to the {\it uncalculable} power-like corrections in the cutoff scale
in the genuine 5D case.
Consequently, we show that the allowed contribution of extra color triplets 
or the breaking scale of the extra $U(1)$ is sensitive to the shape modulus 
in a phenomenologically successful $SO(10)$ model. 

The paper is organized as follows. First we give a brief review on the general boundary 
conditions for breaking the bulk gauge symmetry on $T^2/Z_2$.
In Section 3, we present the one-loop effective action for gauge bosons 
in the general 6D orbifold GUTs and derive the running for the effective
gauge couplings at low energy. Then, in Section 4, we consider the case with
$SO(10)$ bulk group and discuss the gauge coupling unification for some
embeddings of the MSSM. Finally the conclusion is drawn. The details 
on the propagators on GUT orbifolds, the KK summations, the definition of special functions
and some $SO(10)$ group theory facts are given in the appendices.

%%%%%%%%%%%%%%%%%%%%%%%%%%%%%%%%%%%%%%%%%%%%%%%%%%%%%%%%%%%%%%%%%%%%
\section{Boundary conditions on GUT orbifolds}
%%%%%%%%%%%%%%%%%%%%%%%%%%%%%%%%%%%%%%%%%%%%%%%%%%%%%%%%%%%%%%%%%%%%%%%%

Before considering particular models,
we give a brief sketch for the orbifold breaking of gauge symmetry
in a six-dimensional non-Abelian gauge theory with a simple gauge group. 
Two extra dimensions are compactified on the orbifold $T^2/Z_2$. 
For the extra coordinates
$z\equiv x^5+ix^6$, there are double periodicities  
$z\sim z+2\pi (R_5n_5+i R_6n_6)$ 
with radii $R_5, R_6$ and integer numbers $n_5,n_6$. 
Further, when the bulk positions are identified by a $Z_2$ reflection 
symmetry as $z\rightarrow -z$, there are four fixed points on the orbifold: 
$z_0=0$,
$z_1=\pi R_5$, $z_2=i\pi R_6$ and $z_3=\pi R_5+i\pi R_6$.

In order to break the bulk gauge symmetry down to the SM gauge group,
let us introduce nontrivial boundary conditions for bulk gauge 
fields $A_M$ with $M=0,1,2,3\equiv \mu$ and $M=5,6\equiv m$. 
The boundary conditions are specified by unitary
parity matrices $P_i(i=0,1,2,3)$ at the fixed points,
\bea
{\cal P}_i A_\mu(z){\cal P}^{-1}_i
&\equiv& P_i A_\mu(-z+z_i)P_i^{-1}=A_\mu(z+z_i), \nonumber \\
{\cal P}_i A_m(z) {\cal P}^{-1}_i
&\equiv& -P_i A_m(-z+z_i)P_i^{-1}=A_m(z+z_i) \label{bcgauge}
\eea 
where $P^2_i=1(i=0,1,2,3)$. 
The above boundary conditions can be rewritten
simply in terms of component fields with $A_M=A^a_M T_a$ in the group space as
\bea
A^a_\mu(-z+z_i)&=&(Q_i)^a\,_b A^b_\mu(z+z_i), \\
A^a_m(-z+z_i)&=&-(Q_i)^a\,_b A^b_m(z+z_i)
\eea
where 
\be
(Q_i)^a\,_b\equiv {\rm tr}(T^a P_i T_b P_i).
\label{qdef}
\ee
Here the defined matrices $(Q_i)^a\,_b(i=0,1,2,3)$ fulfill 
\be
(Q_i)^a\,_{a'}(Q_i)^b\,_{b'}\eta_{ab}=\eta_{a'b'}, \ \ 
f_{abc}(Q_i)^a\,_{a'}(Q_i)^b\,_{b'}(Q_i)^c\,_{c'}=f_{a'b'c'} \label{qprop}
\ee
where $\eta_{ab}$ is the Killing metric defined 
by ${\rm tr}(T_a T_b)=\eta_{ab}$ on the group space 
and it is used to raise and low adjoint 
indices, and $f_{abc}$ are the group structure constants 
given in the group algebra ${\mbox [T_a,T_b]=if_{abc}T^c}$.
Note that $Q^2_i=1$ from the $Z_2$ symmetry and hence $Q_i$ 
are real symmetric matrices.
Eq.~(\ref{qdef}) and the second property in eq.~(\ref{qprop}) 
can be rewritten, respectively, as  
\bea
P_i T^a P_i&=&(Q_i)^a\,_b T^b, \label{project}\\
Q_i T^a_G Q_i&=&(Q_i)^a\,_b T^b_G 
\eea
with $(T^b_G)^{ac}=if^{abc}$.  

We also discuss on the Wilson lines on a torus in comparison to 
the local boundary conditions as given above.
The boundary conditions along noncontractible loops on a torus are defined
by the unitary matrices $U_1, U_2$ as
\bea
U_1 A_M(z+2\pi R_5)U_1^{-1}&=&A_M(z), \\
U_2 A_M(z+i2\pi R_6) U_2^{-1}&=&A_M(z).
\eea
Since $x^5+\pi R_5\rightarrow -x^5+\pi R_5$ is equivalent to
$x^5+\pi R_5\rightarrow -x^5-\pi R_5\rightarrow -x^5+\pi R_5$
and similarly for the other coordinate,
we obtain the following relations,
\be
U_1=P_1 P_0, \ \ U_2=P_2 P_0.
\ee
Then, we can see that 
the consistency conditions for the Wilson lines on orbifolds,
$U_1 P_0 U_1=P_0$ and $U_2P_0U_2=P_0$, are satisfied.  
Moreover, since $U_2U_1P_0=P_3$ and $[U_1,U_2]=0$, 
the parity matrix $P_3$ can be written as 
\be
P_3=P_2P_0P_1=P_1P_0P_2. \label{p3}
\ee
Therefore, the Wilson lines one can consider are not independent of local
boundary conditions, and one of the parity actions is
not independent.

For simplicity, let us focus on the case with commuting parity matrices,
i.e. ${\mbox [P_i,P_j]=0}$ or $[Q_i,Q_j]=0$. 
For these parity actions, the rank of the gauge group is not reduced.
In this case, it is convenient to choose the Cartan-Weyl basis such that
the orbifold actions become diagonal. 
 In this basis, the generators are
organized into Cartan subalgebra generators $H_I, I=1,\cdots,{\rm rank}(G)$,
and the remaining generators,
$E_\alpha,\alpha=1,\cdots,({\rm dim(G)}-{\rm rank}(G))$, with
\be
[H_I,E_\alpha]=\alpha_IE_\alpha,
\ee
where $\alpha_I$ is the ${\rm rank}(G)-$dimensional
root vector associated with $E_\alpha$.
Then, it is always possible to write the parity matrices as
\be
P_i=e^{-2\pi i V_i\cdot H} \label{paritycartan}
\ee
which defines the ${\rm rank}(G)-$dimensional twist vector $V_i$ for each
fixed point. Thus, the relations (\ref{project}) become
\bea
P_i H_I P_i &=& H_I,  \\
P_i E_\alpha P_i &=& e^{-2\pi i\alpha \cdot V_i}E_\alpha.
\eea
In this basis, the matrices $Q_i$ are also diagonal such that
$(Q_i)^I\,_J=\delta^I_J$ and
$(Q_i)^\alpha\,_\beta=e^{-2\pi i\alpha \cdot V_i}\delta^\alpha_\beta$ and
other entries are zero. 
Here we have that $\alpha \cdot V_i=0$ or $\frac{1}{2}$ mod ${\bf Z}$
for $Z_2$ actions
at the fixed points because $Q^2_i=1$.
Then, a bulk field takes a combination of
parity eigenvalues
$(p_0,p_1,p_2)$ with $p_i=+1$ or $-1$ under three independent $Z_2$ actions,
so it is composed
of a subset of basis functions on a torus with radii $2R_5$ and $2R_6$.

%%%%%%%%%%%%%%%%%%%%%%%%%%%%%%%%%%%%%%%%%%%%%%%%%%%%%%%%%%%%%%%%%%%%
\section{The effective action on GUT orbifolds}
%%%%%%%%%%%%%%%%%%%%%%%%%%%%%%%%%%%%%%%%%%%%%%%%%%%%%%%%%%%%%%%%%%%%%%%%

In this section, we present the general formulae for the one-loop 
effective action in a 6D ${\cal N}=1$ supersymmetric GUT 
where the bulk gauge symmetry is broken by local boundary conditions
at the fixed points on the $T^2/Z_2$ orbifold as described in
the previous section.
As a result, we also discuss about the running of the gauge couplings
for zero-mode gauge bosons at low energy.

\subsection{The one-loop effective action on the $T^2/Z_2$ orbifold}

We consider a 6D ${\cal N}=1$ supersymmetric non-Abelian gauge theory 
compactified on the orbifold $T^2/Z_2$.
In terms of component fields, a vector multiplet is composed 
of gauge bosons $A_M$ and (right-handed) symplectic Majorana 
gauginos $\lambda$  while a hyper multiplet is composed of
two complex hyperscalars $\phi_\pm$ without opposite charges
and a (left-handed) hyperino $\psi$. Since all charged hyperinos
have the equal 6D chiralities due to supersymmetry,
one is not allowed to write the 6D mass terms for hyper multiplets.

In the process of taking the usual 
gauge fixing for a non-Abelian gauge theory\cite{gls}, 
we also introduce ghost fields $c^a$. 
Then, the orbifold 
boundary conditions for bulk component fields 
that we are considering are as the following,
\bea\label{bc}
A^a_\mu(x,-z+z_i)&=&(Q_i)^a\,_b \,A^b_\mu(x,z+z_i), \quad
A^a_m(x,-z+z_i)=-(Q_i)^a\,_b \,A^b_m(x,z+z_i), \nonumber \\
c^a(x,-z+z_i)& =&  (Q_i)^a\,_b \,c^b(x,z+z_i), \quad 
\lambda^a(x,-z+z_i) =i\gamma^5\,(Q_i)^a\,_b\,\lambda^b(x,z+z_i),\nonumber  
\\
\psi(x,-z+z_i)&=&i\gamma^5\,\eta_i P_i\,\psi(x,z+z_i), \nonumber\\
\phi_+(x,-z+z_i)&=& \eta_i P_i\,\phi_+(x,z+z_i), \quad  
\phi_-(x,-z+z_i)= -\eta_i\phi_-(x,z+z_i)P_i \label{bcall}
\eea
with $i=0,1,2,3$.
Here the forms of the parity matrices depend on the representation of 
a hyper multiplet under the bulk gauge group. 
Each hyper multiplet can take its own value of $\eta_i$ 
as either $+1$ or $-1$. 

Taking into account the group structure of propagators in loops
as discussed in the Appendix A 
and following the similar procedure as in the case with 
no orbifold breaking of the gauge symmetry in Ref.~\cite{gls},
we obtain the one-loop effective action for the {\it background}
gauge bosons up to quadratic orders as
\bea
&&\Gamma^{(2)}[A_\mu]=
\!\frac{1}{2g^2}\sum_{{\vec k}}
\int \!\frac{d^4 k}{(2\pi)^4}
A^a_\mu(-k,-{\vec k})\, A^a_\nu(k,{\vec k})
\Big(-(k^2-{\vec k}^2)g^{\mu\nu}+k^\mu k^\nu\Big)
\qquad\qquad\qquad \nonumber \\[7pt]
&&\qquad\qquad
+\frac{i}{2}\sum_{{\vec k},{\vec k}'}
\int \frac{d^4 k}{(2\pi)^4}
A^{b\mu}(-k,-{\vec k}')\, A^\nu_a(k,{\vec k})  \\[5pt]
&& \qquad\qquad\quad\,
 \times  \bigg\{\!  -
\!\Pi_{\mu\nu}^{\rm G}
\!+ \! 4(k^2g_{\mu\nu}\!-\! k_\mu k_\nu)\Pi^G_{++}
-2{\vec k}\cdot{\vec k}'g_{\mu\nu}(\Pi^G_{+-}\!+\!\Pi^G_{-+})-
\Pi_{\mu\nu}^{\rm H} \!\bigg\}^a\,_b\quad \nonumber
\eea
where $\Pi_{\mu\nu}^{\rm G}=\Pi^g_{\mu\nu,+}
+\Pi^g_{\mu\nu,-}+\Pi^\lambda_{\mu\nu}$, 
$\Pi_{\mu\nu}^{\rm H}=\Pi^h_{\mu\nu,+}+\Pi^h_{\mu\nu,-}+\Pi^\psi_{\mu\nu}$
with
\bea
(\Pi^g_{\mu\nu,\pm})^a\,_b&=&\sum_{{\vec p},{\vec p}'}
\int \frac{d^4 p}{(2\pi)^4}{\rm Tr}\bigg[\Big\{
-(2p+k)_\mu(2p+k)_\nu
{\tilde G}_{g,\pm}(p+k,{\vec p}'+{\vec k}',{\vec p}+{\vec k}) \nonumber \\
&&\qquad\qquad\qquad
+2ig_{\mu\nu}\delta_{{\vec p}',{\vec p}+{\vec k}-{\vec k}'}\Big\}T_b
{\tilde G}_{g,\pm}(p,{\vec p},{\vec p}')T^a\bigg], \\
(\Pi^h_{\mu\nu,\pm})^a\,_b&=&\sum_{{\vec p},{\vec p}'}
\int \frac{d^4 p}{(2\pi)^4}{\rm Tr}\bigg[\Big\{
-(2p+k)_\mu(2p+k)_\nu
{\tilde G}_{h,\pm}(p+k,{\vec p}'+{\vec k}',{\vec p}+{\vec k}) \nonumber \\
&&\qquad\qquad\qquad
+2ig_{\mu\nu}\delta_{{\vec p}',{\vec p}+{\vec k}-{\vec k}'}\Big\}T_b
{\tilde G}_{h,\pm}(p,{\vec p},{\vec p}')T^a\bigg], \\
(\Pi^\lambda_{\mu\nu})^a\,_b
&=&\sum_{{\vec p},{\vec p}'}\int \frac{d^4p}{(2\pi)^4}
{\rm Tr}\Big[{\tilde D}_\lambda(p,{\vec p},{\vec p}')\gamma_\mu T_b
{\tilde D}_\lambda(p+k,{\vec p}'+{\vec k}',{\vec p}+{\vec k})
\gamma_\nu T^a\Big], 
\\
(\Pi^\psi_{\mu\nu})^a\,_b&=&\sum_{{\vec p},{\vec p}'}\int \frac{d^4p}{(2\pi)^4}
{\rm Tr}\Big[{\tilde D}_\psi(p,{\vec p},{\vec p}')\gamma_\mu T_b
{\tilde D}_\psi(p+k,{\vec p}'+{\vec k}',{\vec p}+{\vec k})\gamma_\nu T^a\Big],
\eea
and
\bea
(\Pi^G_{\pm\pm})^a\,_b
&=&\sum_{{\vec p},{\vec p}'}\int
\frac{d^4 p}{(2\pi)^4}\,\, {\rm Tr}\bigg[
{\tilde G}_{g,\pm}(p+k,{\vec p}+{\vec k},{\vec p}'+{\vec k}')T_b
{\tilde G}_{g,\pm}(p,{\vec p},{\vec p}')T^a\bigg], \\
(\Pi^G_{\pm\mp})^a\,_b
&=&\sum_{{\vec p},{\vec p}'}\int
\frac{d^4 p}{(2\pi)^4}\,\, {\rm Tr}\bigg[
{\tilde G}_{g,\mp}(p+k,{\vec p}+{\vec k},{\vec p}'+{\vec k}')T_b
{\tilde G}_{g,\pm}(p,{\vec p},{\vec p}')T^a\bigg] \nonumber \\
&=&(\Pi^G_{\pm,\pm})^a\,_b.
\eea
Here the propagators appearing in the loops are given in the Appendix A.
Since we consider the commuting parity matrices, 
the orbifold actions with respect to
the fixed points other that the origin are factorized out 
of the propagators which would be given for the case 
with one $Z_2$ orbifold action only.
Then, after identifying various equivalent terms,
we get the effective action in a simpler form as a decomposition 
into bulk and brane parts, 
\be
\Gamma^{(2)}[A_\mu]=\Gamma_{\rm bulk}+\Gamma_{\rm brane} \label{effaction}
\ee
with
\bea
\Gamma_{\rm bulk}&=&\frac{1}{2}\sum_{{\vec k},{\vec k}'}
\int\frac{d^4 k}{(2\pi)^4}A^b_\mu(-k,-{\vec k})A_{a\nu}(k,{\vec k}')
((k^2-{\vec k}^2)g^{\mu\nu}-k^\mu k^\nu) \nonumber \\
&&\times \bigg[-\frac{1}{g^2}\delta^a_b-i(\Pi_G
+\Pi_H)^a\,_b(k,{\vec k})\bigg]
\delta_{{\vec k},{\vec k}'}, \label{bulk}\\
\Gamma_{\rm brane}&=&\frac{1}{2}\sum_{{\vec k},{\vec k}'}
\int\frac{d^4 k}{(2\pi)^4}A^b_\mu(-k,-{\vec k})A_{a\nu}(k,{\vec k}')
(k^2g^{\mu\nu}-k^\mu k^\nu)[-4i({\tilde\Pi}_G)^a\,_b] \label{brane}
\eea
where
\bea
&&(\Pi_G)^a\,_b(k,{\vec k})=\mu^{4-d}\sum_{\vec p}\int\frac{d^d p}{(2\pi)^d}
\frac{1}{(p^2-{\vec p}^2)[(p+k)^2-({\vec p}+{\vec k})^2]}\nonumber \\
&&\qquad\times\frac{1}{4}{\rm tr}_{\rm Adj}\Big[\Big\{1
+\cos(2p_5\pi R_5)Q_0Q_1\Big\}
\Big\{1+\cos(2p_6\pi R_6)Q_0Q_2\Big\}T^a T_b\Big], \\
&&(\Pi_H)^a\,_b(k,{\vec k})=-\mu^{4-d}\sum_{\vec p}\int\frac{d^d p}{(2\pi)^d}
\frac{1}{(p^2-{\vec p}^2)[(p+k)^2-({\vec p}+{\vec k})^2]} \nonumber \\ 
&&\qquad\times\frac{1}{4}{\rm tr}_{\rm R}\Big[\Big\{1
+\eta_0\eta_1\cos(2p_5\pi R_5)P_0P_1\Big\} 
\Big\{1+\eta_0\eta_2\cos(2p_6\pi R_6) P_0P_2\Big\}
T^a T_b\Big],
\\
&&({\tilde\Pi}_G)^a\,_b(k,{\vec k}',{\vec k})=
\frac{\mu^{4-d}}{2}\sum_{\vec p}\int\frac{d^d p}{(2\pi)^d}
\frac{\delta_{-2{\vec p},{\vec k}-{\vec k}'}}{(p^2-{\vec p}^2)[(p+k)^2-({\vec p}+{\vec k})^2]}\nonumber \\
&&\qquad\times\frac{1}{4}{\rm tr}_{\rm Adj}\Big[\Big\{1+\cos(2p_5\pi R_5)
Q_0Q_1\Big\}
\Big\{1
+\cos(2p_6\pi R_6) Q_0 Q_2\Big\}Q_0T^a T_b\Big].
\eea
Here $\mu$ is the renormalization scale in dimensional regularization
with $d=4-\epsilon$,
and ${\vec p}=(p_5,p_6)=(\frac{n_5}{2R_5},\frac{n_6}{2R_6})$ 
with $n_5,n_6$ being integer,
and similarly for ${\vec k}$ and ${\vec k}'$. In simplifying the expressions
in the above, for the generators satisfying $P_iT^a=\pm T^a P_i$, 
we made use of $\cos(2(p_5+k_5)\pi R_5)=\pm \cos(2p_5 \pi R_5)$
and $\cos(2(p_6+k_6)\pi R_6)=\pm \cos(2p_6 \pi R_6)$.
Further, we notice that ${\rm tr}_{\rm Adj}$ is the trace over indices of the adjoint 
representation and ${\rm tr}_{\rm R}$ is the trace over indices of the 
$R$ representation. 

From eq.~(\ref{effaction}), we can see that a vector multiplet gives rise to both 
bulk and brane-localized corrections while 
a hyper multiplet only leads to a bulk correction. It has been shown that 
the absence of the brane-localized corrections due to a hyper multiplet is restricted
to the case with even ordered orbifolds\cite{nibb}. 

In order to simplify the expression for the bulk contribution (\ref{bulk}),
we define the quantity
\bea
\Pi^{(\rho_5,\rho_6)}&\equiv&\mu^{4-d}\sum_{\vec p}\int\frac{d^d p}{(2\pi)^d}
\frac{1}{(p^2-{\vec p}^2)[(p+k)^2-({\vec p}+{\vec k})^2]} \nonumber \\
&=&\frac{i}{(4\pi)^2V}(2\pi\mu)^\epsilon
\int^1_0 dx\, {\cal J}_0[x(1-x)(k^2+{\vec k}^2),xk_5R_5+\rho_5,xk_6 R_6+\rho_6]
\label{pi_bulk}
\eea
where
${\vec p}=(\frac{n_5+\rho_5}{R_5},\frac{n_6+\rho_6}{R_6})$
with $\rho_5,\rho_6=0$  or $\frac{1}{2}$ and $n_5,n_6$ being integer,
$V\equiv (2\pi)^2 R_5 R_6$, and
\be
{\cal J}_0[c,c_1,c_2]\equiv \sum_{n_1,n_2\in {\bf Z}}\int^\infty_0
\frac{dt}{t^{1-\epsilon/2}}\,e^{-\pi t[c+a_1(n_1+c_1)^2+a_2(n_2+c_2)^2]}
\ee
with $a_i=1/R^2_{i+4} (i=1,2)$. 
Thus, we can rewrite the bulk contribution due to a vector multiplet
as
\bea
(\Pi_G)^a\,_b&=&\frac{1}{4}{\rm tr}_{\rm Adj}[T^a T_b]\,
(\Pi^{(0,0)}+\Pi^{(0,\frac{1}{2})}+\Pi^{(\frac{1}{2},0)}
+\Pi^{(\frac{1}{2},\frac{1}{2})})
\nonumber \\
&&+\frac{1}{4}{\rm tr}_{\rm Adj}[Q_0Q_1T^a T_b]\,
(\Pi^{(0,0)}
+\Pi^{(0,\frac{1}{2})}-\Pi^{(\frac{1}{2},0)}
-\Pi^{(\frac{1}{2},\frac{1}{2})}) \nonumber \\
&&+\frac{1}{4}{\rm tr}_{\rm Adj}[Q_0Q_2T^a T_b]\,
(\Pi^{(0,0)}
-\Pi^{(0,\frac{1}{2})}+\Pi^{(\frac{1}{2},0)}
-\Pi^{(\frac{1}{2},\frac{1}{2})}) \nonumber \\
&&+\frac{1}{4}{\rm tr}_{\rm Adj}[Q_1Q_2T^a T_b]\,
(\Pi^{(0,0)}
-\Pi^{(0,\frac{1}{2})}-\Pi^{(\frac{1}{2},0)}
+\Pi^{(\frac{1}{2},\frac{1}{2})}). \label{vectorpart}
\eea
Similarly, we can write the bulk contribution of a hyper multiplet as
\bea
(\Pi_H)^a\,_b&=&-\frac{1}{4}{\rm tr}_{\rm R}[T^a T_b]\,(\Pi^{(0,0)}
+\Pi^{(0,\frac{1}{2})}+\Pi^{(\frac{1}{2},0)}
+\Pi^{(\frac{1}{2},\frac{1}{2})})
\nonumber \\
&&-\frac{1}{4}\eta_0\eta_1\,{\rm tr}_{\rm R}[P_0P_1T^a T_b]\,(\Pi^{(0,0)}
+\Pi^{(0,\frac{1}{2})}-\Pi^{(\frac{1}{2},0)}
-\Pi^{(\frac{1}{2},\frac{1}{2})}) \nonumber \\
&&-\frac{1}{4}\eta_0\eta_2{\rm tr}_{\rm R}[P_0P_2T^a T_b]\,(\Pi^{(0,0)}
-\Pi^{(0,\frac{1}{2})}+\Pi^{(\frac{1}{2},0)}
-\Pi^{(\frac{1}{2},\frac{1}{2})}) \nonumber \\
&&-\frac{1}{4}\eta_1\eta_2
{\rm tr}_{\rm R}[P_1P_2T^a T_b]\,(\Pi^{(0,0)}
-\Pi^{(0,\frac{1}{2})}-\Pi^{(\frac{1}{2},0)}
+\Pi^{(\frac{1}{2},\frac{1}{2})}).\label{hyperpart}
\eea

Also defining 
\bea
{\tilde\Pi}^{(\rho_5,\rho_6)}&\equiv&
\frac{\mu^{4-d}}{2}\sum_{\vec p}\int \frac{d^d p}{(2\pi)^d}
\frac{\delta_{-2{\vec p},{\vec k}-{\vec k}'}}
{(p^2-{\vec p}^2)[(p+k)^2-({\vec p}+{\vec k})^2]}
\eea
where
${\vec p}=(\frac{n_5+\rho_5}{R_5},\frac{n_6+\rho_6}{R_6})$ 
with
$\rho_5,\rho_6=0$ or $\frac{1}{2}$ and $n_5,n_6$ being integer, 
we can rewrite the brane contribution as
\bea
({\tilde \Pi}_G)^a\,_b&=&\frac{1}{4}{\rm tr}_{\rm Adj}[Q_0T^a T_b] 
\,({\tilde\Pi}^{(0,0)}+{\tilde\Pi}^{(0,\frac{1}{2})}
+{\tilde\Pi}^{(\frac{1}{2},0)}
+{\tilde\Pi}^{(\frac{1}{2},\frac{1}{2})})
\nonumber \\
&&+\frac{1}{4}{\rm tr}_{\rm Adj}[Q_1T^a T_b]\,({\tilde\Pi}^{(0,0)}
+{\tilde\Pi}^{(0,\frac{1}{2})}-{\tilde\Pi}^{(\frac{1}{2},0)}
-{\tilde\Pi}^{(\frac{1}{2},\frac{1}{2})}) \nonumber \\
&&+\frac{1}{4}{\rm tr}_{\rm Adj}[Q_2T^a T_b]\,({\tilde\Pi}^{(0,0)}
-{\tilde\Pi}^{(0,\frac{1}{2})}+{\tilde\Pi}^{(\frac{1}{2},0)}
-{\tilde\Pi}^{(\frac{1}{2},\frac{1}{2})}) \nonumber \\
&&+\frac{1}{4}{\rm tr}_{\rm Adj}[Q_3T^a T_b]\,({\tilde\Pi}^{(0,0)}
-{\tilde\Pi}^{(0,\frac{1}{2})}-{\tilde\Pi}^{(\frac{1}{2},0)}
+{\tilde\Pi}^{(\frac{1}{2},\frac{1}{2})}).\label{vectorbpart}
\eea  
Thus, we find that the brane-localized contribution at each fixed point 
corresponds to the brane projection of the bulk quantity 
by the local parity matrix.

\subsection{Counterterms for loop divergences}

After the KK summation given in the Appendix B, we can separate the divergent
term as 
\be
\Pi^{(\rho_5,\rho_6)}=\frac{i}{(4\pi)^2V}\frac{\pi}{6}
R_5 R_6(k^2+{\vec k}^2)\bigg[\frac{-2}{\epsilon}\bigg]+{\cal O}(\epsilon^0).
\ee
Thus, the bulk contribution becomes
\bea
(\Pi_G+\Pi_H)^a\,_b=\frac{1}{4}({\rm tr}_{\rm Adj}[T^a T_b]-
{\rm tr}_{\rm R}[T^aT_b])\frac{i}{(4\pi)^2V}\frac{\pi}{6}
R_5 R_6(k^2+{\vec k}^2)\bigg[\frac{-2}{\epsilon}\bigg]+{\cal O}(\epsilon^0).
\label{vectorpole}
\eea
Therefore, we find that the loop corrections generate a divergent higher 
derivative term the coefficient of which is proportional 
to the universal ${\cal N}=2$ beta function. 
At the momentum scale higher than the compactification scales, 
the higher derivative operator becomes important so that 
the gauge couplings run power-like in momentum scale rather than logarithmically\cite{gls}. 
However, it does not affect the unification of the gauge couplings, even if 
it is important in determining 
the value of the unified gauge coupling and the unification scale. 

For a given set of ingoing and outgoing momenta of gauge bosons satisfying
${\vec p}=\frac{{\vec k}'-{\vec k}}{2}$,
we compute ${\tilde\Pi}^{(\rho_5,\rho_6)}$ as
\be
{\tilde\Pi}^{(\rho_5,\rho_6)}=\frac{i}{32\pi^2}
\bigg\{\frac{2}{\epsilon}+\ln(4\pi\mu^2 e^{-\gamma_E})
-\int^1_0 dx\,\ln\bigg[x(1-x)(k^2+{\vec k}^2)+\Big(\frac{{\vec k}'}{2}
+(x-\frac{1}{2}){\vec k}\Big)^2\bigg]\bigg\}.\label{vectorbcomp}
\ee
Thus, the brane contribution becomes
\bea
({\tilde \Pi}_G)^a\,_b=\frac{1}{4}\frac{i}{32\pi^2}
\frac{2}{\epsilon}\sum_{i=0}^3{\rm tr}_{\rm Adj}[Q_iT^a T_b]
\,\delta^2(z-z_i)
+{\cal O}(\epsilon^0).\label{vectorbpole}
\eea
Therefore, the appearing divergent term at each fixed
point respects the corresponding gauge symmetry which depends on the local orbifold action.

Consequently, in order to subtract the $\epsilon$ poles 
in $(\Pi_G)^a\,_b, (\Pi_H)^a\,_b$ and $({\tilde\Pi}_G)^a\,_b$
obtained in eqs.~(\ref{vectorpole}) and (\ref{vectorbpole}),
we require the following new counterterms which are not present in the original
action,
\bea
{\cal L}_{c.t.}=\int d^2z d^2\theta\bigg[\frac{1}{2h^2}
{\rm tr}[W\Box_6 W]
+\frac{1}{2}\sum_{i=0}^3
\bigg(\frac{1}{g^2_{i,a}} W_a W^a
\delta^2(z-z_i)\bigg)\bigg]
+{\rm h.c.}
\eea
Here $h^2$ is a dimensionless bulk coupling while
$g^2_{i,a}(i=0,1,2,3)$ are dimensionless gauge couplings correspoding to
the local gauge groups at the fixed points.
Note that the brane-localized gauge coupling 
can be non-universal
so it could affect the predictive power of the orbifold GUTs.

\subsection{Limiting cases}

For a later use in the running of the zero-mode gauge coupling, 
let us take the asymptotic limits of the loop corrections.
First, in the low momentum limit $k^2\ll 1/R^2_{5,6}$, with ${\vec k}={\vec k}'=0$,
by using eq.~(\ref{limit-j0}),
we get the approximate forms for eq.~(\ref{pi_bulk}),
\bea
\Pi^{(0,0)}&\approx&\frac{i}{(4\pi)^2V}\bigg\{\frac{\pi}{6}R_5R_6k^2
\bigg[\frac{-2}{\epsilon}-\ln\Big[\pi e^{\gamma_E}\mu^2R^2_5|\eta(iu)|^{-4}
\Big]\bigg] \nonumber \\
&&-\ln\Big[4\pi^2 e^{-2}|\eta(iu)|^4 R^2_6 k^2\Big]\bigg\}, \label{limit1}\\
\Pi^{(0,\frac{1}{2})}&\approx&\frac{i}{(4\pi)^2V}\bigg\{\frac{\pi}{6}R_5R_6k^2
\bigg[\frac{-2}{\epsilon}-\ln(\pi e^{\gamma_E}\mu^2R^2_5)-\pi u
+2\sum_{n\geq 1}(-1)^n\ln|1-e^{-2\pi u n}|^2\bigg] \nonumber \\
&&-\ln\bigg|\frac{\vartheta_1(1/2|iu)}{\eta(iu)}\bigg|^2\bigg\}, 
\label{limit2}\\
\Pi^{(\frac{1}{2},0)}&\approx&\frac{i}{(4\pi)^2V}\bigg\{\frac{\pi}{6}R_5R_6k^2
\bigg[\frac{-2}{\epsilon}-\ln(4\pi e^{\gamma_E}\mu^2R^2_5)
-2\sum_{n\geq 1}
\ln\bigg|\frac{1+e^{-\pi un}}{1-e^{-\pi un}}\bigg|^2\bigg] \nonumber \\
&&-\ln\bigg\vert\frac{\vartheta_1(-i u/2\vert i u)}{\eta(i u) }e^{-\pi u/4}\bigg\vert^2
\bigg\}, \label{limit3}\\
\Pi^{(\frac{1}{2},\frac{1}{2})}&\approx&\frac{i}{(4\pi)^2V}\bigg\{\frac{\pi}{6}R_5R_6k^2
\bigg[\frac{-2}{\epsilon}-\ln(4\pi e^{\gamma_E}\mu^2R^2_5)
-2\sum_{n\geq 1}(-1)^n
\ln\bigg|\frac{1+e^{-\pi un}}{1-e^{-\pi un}}\bigg|^2\bigg] \nonumber \\
&&-\ln\bigg\vert\frac{\vartheta_1(1/2-i u/2\vert i u)}{\eta(iu)} e^{-\pi u/4}\bigg\vert^2
\bigg\}.\label{limit4}
\eea
Further, using the fact that $\Pi^{(0,0)}+\Pi^{(0,\frac{1}{2})}+
\Pi^{(\frac{1}{2},0)}+\Pi^{(\frac{1}{2},\frac{1}{2})}$ is the same
as the KK sum on a torus with each radius double sized and with no Wilson lines, 
i.e. from the approximate form of the sum, 
\bea
\sum_{\rho_5,\rho_6=0,\frac{1}{2}}\Pi^{(\rho_5,\rho_6)}&\approx&\frac{i}{(4\pi)^2V}
\bigg\{\frac{\pi}{6}(4R_5R_6)k^2
\bigg[\frac{-2}{\epsilon}-\ln\Big[4\pi e^{\gamma_E}\mu^2R^2_5|\eta(iu)|^{-4}
\Big]\bigg] \nonumber \\
&&-\ln\Big[16\pi^2 e^{-2}|\eta(iu)|^4 R^2_6 k^2\Big]\bigg\},
\eea
we note the useful identity for the theta functions,
\be
\Big|\vartheta_1(1/2|iu)\,\vartheta_1(-i u/2\vert i u)\,
\vartheta_1(1/2-i u/2\vert i u)\Big|^2= 4|\eta(iu)|^6 e^{\pi u}.\label{identity}
\ee

In the high momentum limit $k^2\gg 1/R^2_{5,6}$, with ${\vec k}={\vec k}'=0$,
eq.~(\ref{pi_bulk}) becomes, independently of the orbifold actions,
\bea
\Pi^{(\rho_5,\rho_6)}
\approx \frac{i}{(4\pi)^2V}\bigg\{\frac{\pi}{6}R_5R_6k^2
\bigg[\frac{-2}{\epsilon}-\ln\frac{\mu^2}{k^2}
-\ln\Big(4\pi e^{8/3-\gamma_E}\Big)\bigg]\bigg\}.
\eea
Therefore, from eqs.~(\ref{vectorpart}) and (\ref{hyperpart}), 
even the finite part of the bulk correction
becomes universal at high energy.

\subsection{Running of the 4D effective gauge coupling} 

In this section, we consider the running of 
the effective gauge coupling which is defined as the coefficient
of the kinetic term of a zero-mode gauge boson. It also includes the bulk
higher derivative term and the brane kinetic terms.

From the one-loop effective action (\ref{effaction}), 
the zero-mode gauge coupling reads 
\bea
\frac{1}{g^2_{{\rm eff},ab}(k^2)}
=\frac{1}{g^2_{{\rm tree},ab}}-\frac{k^2 V}{h^2_{\rm tree}}\,
\delta_{ab}+iV(\Pi_G(k,0)+\Pi_H(k,0))^a\,_b+4i({\tilde\Pi}_G(k,0,0))^a\,_b
\label{grun00}
\eea
with
\be
\frac{1}{g^2_{{\rm tree},ab}}=\bigg[\frac{V}{g^2}
+\sum_{i=0}^3\frac{1}{g^2_{i,a}}\bigg]\delta_{ab}.
\ee
When taking the minimal subtraction scheme for divergences (\ref{vectorpole}) 
and (\ref{vectorbpole}) at $k^2=M^2_*$, where $M_*$ is the 6D fundamental scale,
we define the renormalized bulk and brane gauge couplings for $\xi_1\mu^2=M^2_*$
with $\xi_1=4\pi e^{2-\gamma_E}$ at that scale.
Then, below the compactification scales ($k^2\ll 1/R^2_{5,6}$),
using eqs.~(\ref{vectorbcomp}), (\ref{limit1})-(\ref{limit4}) 
with (\ref{identity}),
we have eq.~(\ref{grun00}) as
\bea
\frac{1}{g^2_{{\rm eff},ab}(k^2)}
=\frac{1}{g^2_{{\rm r},a}}\delta_{ab}
+\frac{1}{16\pi^2}B_{ab}\ln\frac{M^2_*}{k^2} 
-\frac{1}{16\pi^2}\sum_{i,j=\pm}B^{ij}_{ab}L_{ij}
-\frac{1}{4\pi}\kappa_{ab}
\label{generalrun}
\eea
where $g_{{\rm r},a}$ are the renormalized gauge couplings and
the beta functions are 
\bea
B_{ab}&=& \frac{1}{4}\bigg[-{\rm tr}_{\rm Adj}(T^aT_b)+\sum_R{\rm tr}_{\rm R}(T^aT_b) 
-2\sum_{i=0}^3{\rm tr}_{\rm Adj}(Q_iT^aT_b)\bigg] \nonumber \\
&&+\frac{1}{4}\sum_{i=1}^3\Big[-{\rm tr}_{\rm Adj}(Q_0Q_iT^aT_b)
+\sum_R\eta^R_0\eta^R_i{\rm tr}_{\rm R}(P_0P_iT^aT_b)\Big]
\eea
and
\bea
B^{++}_{ab}&=&\frac{1}{4}\bigg[-{\rm tr}_{\rm Adj}(T^aT_b)
+\sum_R{\rm tr}_{\rm R}(T^aT_b)\bigg], \\
B^{-+}_{ab}&=&\frac{1}{4}\bigg[-{\rm tr}_{\rm Adj}(Q_0Q_1T^aT_b)
+\sum_R\eta^R_0\eta^R_1{\rm tr}_{\rm R}(P_0P_1T^aT_b)\bigg], \\
B^{+-}_{ab}&=&\frac{1}{4}\bigg[-{\rm tr}_{\rm Adj}(Q_0Q_2T^aT_b)
+\sum_R\eta^R_0\eta^R_2{\rm tr}_{\rm R}(P_0P_2T^aT_b)\bigg], \\
B^{--}_{ab}&=&\frac{1}{4}\bigg[-{\rm tr}_{\rm Adj}(Q_0Q_3T^aT_b)
+\sum_R\eta^R_0\eta^R_3{\rm tr}_{\rm R}(P_0P_3T^aT_b)\bigg]
\eea
with
\bea
L_{++}&=&\ln\Big[4e^{-2}|\eta(iu)|^4uVM^2_*\Big], \\
L_{-+}&=&\ln\Big[\frac{e^{-2}}{4}\Big|\vartheta_1(\frac{1}{2}|iu)\Big|^4uV
M^2_*\Big], \\
L_{+-}&=&\ln\Big[\frac{e^{-2}}{4}\Big|\vartheta_1
(-\frac{1}{2}iu|iu)e^{-\pi u/4}\Big|^4uVM^2_*\Big], \\
L_{--}&=&\ln\Big[\frac{e^{-2}}{4}
\Big|\vartheta_1(\frac{1}{2}-\frac{1}{2}iu|iu)e^{-\pi u/4}\Big|^4 uVM^2_*\Big].
\eea
Further, $\kappa_{ab}$ corresponds to the power-like dependence on the 
momentum scale and it is suppressed by the compactification volume 
at low energy
as in the case without orbifold breaking of the gauge symmetry\cite{gls}.
$B_{ab}$ are the ${\cal N}=1$ beta function coefficients of the logarithmic
running due to the massless modes. 
They are composed of both bulk and brane corrections.
On the other hand, $B^{ij}_{ab}$ 
are the ${\cal N}=2$ beta function coefficients for the KK massive modes
of the bulk fields. 
The logarithms $L_{ij}$ have the common volume ($V$) dependence, but also they
are functions of the shape modulus ($u$), being of different form 
depending on the parities.
Since the KK massive mode correction contains 
a non-universal part due to the gauge symmetry breaking, 
they can affect the unification of gauge couplings.

\section{Gauge coupling unification in a 6D $SO(10)$ orbifold GUT}

We consider a 6D ${\cal N}=1$ supersymmetric $SO(10)$ orbifold GUT and  
compute the gauge coupling running by using the general
formulae found in the previous section.

\subsection{Orbifold breaking of $SO(10)$}

In order to break the bulk $SO(10)$ gauge group down to the SM one,
we introduce the parity matrices in eq.~(\ref{bcgauge}) or (\ref{bcall})
for a fundamental representation \cite{abc} as
\bea
P_0&=&I_{10\times 10}, \\ 
P_1 &=&  {\rm diag}(-1,-1,-1,1,1)\times \sigma^0, \\
%\left(\begin{array}{lllll}  -\sigma^0 & & & &  \\ 
% &-\sigma^0 & & & \\  & & -\sigma^0 & & \\ 
%& &  & \sigma^0 & \\ & &  & & \sigma^0\end{array}\right), \\
P_2&=&  {\rm diag}(1,1,1,1,1)\times \sigma^2,
%\left(\begin{array}{lllll} \,\,\,\sigma^2 & & & &  \\ 
% &\,\,\,\sigma^2 & & & \\  & & \,\,\,\sigma^2 & & \\ 
%& &  & \,\,\,\sigma^2 & \\ & &  & & \,\,\,\sigma^2\end{array} \right)
\eea
and $P_3=P_1P_2$ from the consistency condition (\ref{p3}).
Then, 
the parity operations $P_1, P_2$ break $SO(10)$ down to
maximal subgroups, the Pati-Salam group $SU(4)\times SU(2)_L\times SU(2)_R$ and 
the Georgi-Glashow group $SU(5)\times U(1)_X$, respectively. 
The parity operation $P_3$ also breaks $SO(10)$ down to the flipped $SU(5)$ but it
is not an independent breaking. 
Thus, the intersection of two maximal subgroups 
leads to $SU(3)_C\times SU(2)_L\times U(1)_Y\times U(1)_X$ 
as the remaining gauge group. This can be seen from the gauge bosons with
positive parities:
${\bf 45}$ is decomposed into $\bf (15,1,1)_++(6,2,2)_-+(1,3,1)_++(1,1,3)_+$
under $P_1$ (where $\pm$ indicate the parities)
and $\bf 24_{0,+}+10_{-4,-}+\overline {10}_{4,-}+1_{0,+}$ under $P_2$.
Then, finally, the extra $U(1)_X$ or $U(1)_{B-L}$
has to be broken further by the VEV of bulk or brane Higgs fields.

For applying the parity action to other representations, 
from eq.~(\ref{paritycartan}), we can rewrite the parity matrices 
in terms of Cartan-Weyl generators\footnote{One has to be careful with multiplying 
a $U(1)$ phase for the correct parity matrices satisfying $P^2_i=1$.} 
as a special case of eq.~(\ref{paritycartan}),
\bea
P_1&=&e^{-2\pi i x_1(-6T_Y+T_X)},  \ \ x_1=\frac{1}{2}, \\
P_2&=&e^{-2\pi i x_2 T_X}, \ \ x_2=\frac{1}{8} 
\eea
where $T_Y, T_X$ are the $U(1)_Y$ 
and $U(1)_X$ generators\footnote{See the appendix D 
for details.}, respectively.
We consider
a set of hyper multiplets, $N_{10}$ $\bf 10$'s and $N_{16}$ $\bf 16$'s
satisfying $N_{10}=2+N_{16}$ for no irreducible
anomalies\cite{6danomaly,anomaly6dlocal}.
Both $N_{10}$ and $N_{16}$ have to be even for the absence of localized
anomalies unless there are split multiplets
at the fixed points\cite{anomaly6dlocal}.
A ${\bf 10}=(H,G,H^c,G^c)$ is 
decomposed into $\bf (6,1,1)_-+(1,2,2)_+$ under $P_1$
and $\bf 5_{-2,-}+{\bar 5}_{2,+}$ under $P_2$. Then, we get a massless Higgs doublet
from $H^c$ of $\bf 10$.
On the other hand, a ${\bf 16}=(Q,L,U,E,D^c,N^c)$ is
decomposed into $\bf (4,2,1)_++({\bar 4},1,2)_-$ under $P_1$
and $\bf 10_{1,-}+{\bar 5}_{-3,+}+1_{5,+}$ under $P_2$. 
Then, we also get a massless lepton doublet from $L$ of $\bf 16$.
The ${\cal N}=2$ partner of
each hyper multiplet has the parity matrices
in the group space multiplied by the negative overall parity, due to the discrete choice of
the Scherk-Schwarz twist in $SU(2)_R$ space.
We note that in eq.~(\ref{bcall}), $\eta_0=1$ and $\eta_3=\eta_1\eta_2$ for each hyper multiplet.

\subsection{Gauge coupling running at low energy}

In order to break the extra $U(1)_X$ gauge symmetry 
by a usual Higgs mechanism,
one can introduce $\bf 16$ Higgs multiplets in the bulk\cite{abc42,abcflavor}. 
In that case, after the orbifolding, on top of SM singlets, 
one ends up with extra color triplets as zero modes.
Since the extra color triplets can get masses of order the $B-L$
breaking scale $M_{B-L}$ at the fixed points,
they already start contributing to the running of gauge couplings 
at that scale.
Thus,  from the general result (\ref{generalrun}),
we consider the logarithmic running due to zero modes 
by taking two steps across the $B-L$ breaking scale. 

The brane-localized $B-L$ breaking masses for the color triplets 
can modify their KK massive modes so that there exists an additional
contribution to the the gauge coupling running.
However, when the $B-L$ breaking scale is below the compactification scale,
the new contribution becomes suppressed as $M^2_{B-L}/M^2_c$ 
with $M_c\equiv 1/\sqrt{V}$.
Thus, henceforth 
we assume this case to ignore the effect of the brane-localized 
$B-L$ breaking masses.
Then, much below the compactification scale, 
the running of the 4D effective gauge coupling of the SM gauge group 
is governed by
\bea
\frac{1}{g^2_{{\rm eff}, ab}(k^2)}&=&
\frac{1}{g^2_u}\delta_{ab}
+\frac{1}{16\pi^2}B'_{ab}\ln\frac{M^2_{B-L}}{k^2} 
+\frac{1}{16\pi^2}{\tilde B}_{ab}\ln\frac{M^2_*}{M^2_{B-L}} \nonumber \\ 
&&-\frac{1}{16\pi^2}\sum_{i,j=\pm}B^{ij}_{ab}L_{ij}
+\frac{1}{8\pi^2}\Delta_a\delta_{ab} \label{grun0}
\eea
where $g_u$ is the universal renormalized gauge coupling\footnote{Although there are also
power-like threshold corrections in the cutoff regularization\cite{dienes,gls},
they don't contribute to the differential running of gauge couplings.
Nevertheless, the power-like contributions
may have the net effect of placing an upper limit on the possible volume of
the extra dimensions\cite{dienes2}.} 
and $\Delta_a$ are corrections due to
renormalized gauge couplings localized
at the Pati-Salam and flipped $SU(5)$ fixed points.
$B'_{ab}=b_a \delta_{ab}$ are the beta functions 
of the gauge couplings 
in the MSSM with $b_a=(33/5,1,-3)$ given below the $B-L$ breaking scale.
Moreover, above the $B-L$ breaking scale, the beta functions
for the gauge couplings 
are given by ${\tilde B}_{ab}=B_{ab}-C_{ab}+{\hat B}_{ab}$ 
where
\bea
B_{ab}&=& \frac{1}{4}\Big[-3{\rm tr}_{\rm Adj}(T^aT_b)
+\sum_R{\rm tr}_{\rm R}(T^aT_b)\Big] \nonumber \\
&&+\frac{1}{4}\sum_{i=1}^3\Big[-3{\rm tr}_{\rm Adj}(Q_iT^aT_b)
+\sum_R\eta^R_i{\rm tr}_{\rm R}(P_iT^aT_b)\Big] 
\eea
with $Q_3=Q_1Q_2$, $P_3=P_1P_2$ and $\eta^R_3=\eta^R_1\eta^R_2$,
and $C_{ab}=c_a\delta_{ab}$ is the contribution coming from vector-like 
massless modes which get tree-level brane masses of order the GUT scale
and ${\hat B}_{ab}={\hat b}_a \delta_{ab}$ comes from the brane-localized fields.
Further, the beta functions of the KK massive mode corrections are
\bea
B^{++}_{ab}&=&\frac{1}{4}\bigg[-{\rm tr}_{\rm Adj}(T^aT_b)
+\sum_R{\rm tr}_{\rm R}(T^aT_b)\bigg], \\
B^{-+}_{ab}&=&\frac{1}{4}\bigg[-{\rm tr}_{\rm Adj}(Q_1T^aT_b)
+\sum_R\eta^R_1{\rm tr}_{\rm R}(P_1T^aT_b)\bigg], \\
B^{+-}_{ab}&=&\frac{1}{4}\bigg[-{\rm tr}_{\rm Adj}(Q_2T^aT_b)
+\sum_R\eta^R_2{\rm tr}_{\rm R}(P_2T^aT_b)\bigg], \\
B^{--}_{ab}&=&\frac{1}{4}\bigg[-{\rm tr}_{\rm Adj}(Q_3T^aT_b)
+\sum_R\eta^R_3{\rm tr}_{\rm R}(P_3T^aT_b)\bigg].
\eea

\subsubsection{Computation of traces}

Now we compute the necessary traces to get the running equations. 
To this, we define the following invariant quantity 
including all $SO(10)$ gauge fields, 
\bea
B&\equiv& B_{ab}F^a_{SO(10)}F^b_{SO(10)} \nonumber \\
&\equiv& B_V+B_M
\eea
where 
\bea
B_V&\equiv&-\frac{3}{4}\Big[{\rm tr}_{\rm Adj}F^2_{SO(10)}
+\sum_{i=1}^3{\rm tr}_{\rm Adj}(Q_iF^2_{SO(10)})\Big], \\
B_M&\equiv&\sum_R\frac{1}{4}
\Big[{\rm tr}_R F^2_{SO(10)}
+\sum_{i=1}^3\eta^R_i{\rm tr}_R(P_iF^2_{SO(10)})\Big].
\eea
Moreover, similarly we define 
\bea
B^{ij}\equiv B^{ij}_{ab}F^a_{SO(10)}F^b_{SO(10)}. 
\eea

By using the traces for maximal subgroups of $SO(10)$ in the Appendix D,
we obtain the following result for the vector multiplet,  
\bea
B_V
&=&-\frac{3}{4}{\rm tr}_{\rm Adj}F^2_{SO(10)}+6{\rm tr}_2F^2_{SU(2)_L}
+6{\rm tr}_2F^2_{SU(2)_R} \nonumber \\
&&-3{\rm tr}_5 F^2_{SU(5)}+6 F^2_{U(1)_X}
-3{\rm tr}_5 F^2_{SU(5)'}+6 F^2_{U(1)_{X'}}. \label{vectorbeta}
\eea
For the hyper multiplets, we also have 
\be
B_M=B_{10}+B_{16} 
\ee
with
\bea
B_{10}
&=&\frac{1}{4}N_{10}{\rm tr}_{10}F^2_{SO(10)}
+\frac{1}{2}\sum_{10}
\eta^{10}_1\Big[-{\rm tr}_4F^2_{SU(4)}+{\rm tr}_2F^2_{SU(2)_L}
+{\rm tr}_2F^2_{SU(2)_R}\Big], \label{cash10}\\
B_{16}
&=&\frac{1}{4}N_{16}{\rm tr}_{16}F^2_{SO(10)}
+\sum_{16}\eta^{16}_1\Big[{\rm tr}_2F^2_{SU(2)_L}-{\rm tr}_2F^2_{SU(2)_R}\Big] 
\nonumber \\
&&+\frac{1}{4}\sum_{16}\eta^{16}_2\Big[-2{\rm tr}_5 F^2_{SU(5)}
+\frac{3}{2}F^2_{U(1)_X}\Big] \nonumber \\
&&+\frac{1}{4}\sum_{16}\eta^{16}_1\eta^{16}_2\Big[-2{\rm tr}_5 F^2_{SU(5)'}
+\frac{3}{2}F^2_{U(1)_{X'}}\Big] \label{cash16}
\eea
Further, we have
\bea
B^{++} 
&=&\frac{1}{4}[-{\rm tr}_{\rm Adj}F^2_{SO(10)}
+N_{10}{\rm tr}_{10}F^2_{SO(10)}+N_{16}{\rm tr}_{16}F^2_{SO(10)}], \\
B^{-+}
&=&2{\rm tr}_2F^2_{SU(2)_L}+2{\rm tr}_2F^2_{SU(2)_R} \nonumber \\
&&+\frac{1}{2}\sum_{10}\eta^{10}_1\Big[-{\rm tr}_4F^2_{SU(4)}
+{\rm tr}_2F^2_{SU(2)_L}
+{\rm tr}_2F^2_{SU(2)_R}\Big] \nonumber \\
&&+\sum_{16}\eta^{16}_1\Big[{\rm tr}_2F^2_{SU(2)_L}
-{\rm tr}_2F^2_{SU(2)_R}\Big],\\
B^{+-}
&=&-{\rm tr}_5 F^2_{SU(5)}+2F^2_{U(1)_X} \nonumber \\
&&+\frac{1}{4}\sum_{16}\eta^{16}_2\Big[-2{\rm tr}_5 F^2_{SU(5)}
+\frac{3}{2}F^2_{U(1)_X}\Big] \\
B^{--}
&=&-{\rm tr}_5 F^2_{SU(5)'}+2F^2_{U(1)_{X'}} \nonumber \\
&&+\frac{1}{4}\sum_{16}\eta^{16}_1\eta^{16}_2\Big[-2{\rm tr}_5 F^2_{SU(5)'}
+\frac{3}{2}F^2_{U(1)_{X'}}\Big]. 
\eea
Therefore, by reading off the gauge kinetic terms for the SM gauge group 
in $B$ and 
$B^{ij}$, we can find the general expression 
for the beta function coming from the massless modes in the bulk as
$B_{ab}=b_a \delta_{ab}$ with
\be
b_a=b^V_a+b^{10}_a+b^{16}_a \label{bmassless}
\ee
where
\bea
b^V_a&=&(0,-6,-9), \\
b^{10}_a&=&\frac{1}{4}N_{10}(1,1,1)
+\frac{1}{4}\sum_{10}\eta^{10}_1(\frac{1}{5}, 1,-1), \label{b10}
\\
b^{16}_a&=&\frac{1}{4}(2N_{16}-\sum_{16}\eta^{16}_2)(1,1,1)
+\frac{1}{4}\sum_{16}\eta^{16}_1(-\frac{6}{5},2,0)
\nonumber \\
&&+\frac{1}{4} \sum_{16}\eta^{16}_1\eta^{16}_2(\frac{7}{5},-1,-1),
\label{b16}
\eea
and the beta function for KK massive modes 
as $B^{ij}_{ab}=b^{ij}_a \delta_{ab}$ with
\bea
b^{++}_a&=&\frac{1}{4}(-8+N_{10}+2N_{16})(1,1,1), \label{b++}\\
b^{-+}_a&=&\frac{1}{4}(\frac{12}{5},4,0)+\frac{1}{4}
\sum_{10}\eta^{10}_1(\frac{1}{5},1,-1)
+\frac{1}{4}\sum_{16}\eta^{16}_1(-\frac{6}{5},2,0), \label{b-+}\\
b^{+-}_a&=&\frac{1}{4}(2+\sum_{16}\eta^{16}_2)(-1,-1,-1), \label{b+-}\\
b^{--}_a&=&\frac{1}{4}(\frac{38}{5},-2,-2)
+\frac{1}{4}\sum_{16}\eta^{16}_1\eta^{16}_2(\frac{7}{5},-1,-1).\label{b--}
\eea
Here, in order to get the beta function for $U(1)_Y$,  
we made use of the relations between $U(1)$ gauge bosons (\ref{u1grel}) 
in the Appendix D. The corrections due to hyper multiplets in  
eq.~(\ref{cash10}) and (\ref{cash16})
also contain mixing terms\footnote{The
gauge kinetic terms localized 
at Pati-Salam and flipped $SU(5)$ fixed points can also
lead to a mixing.} between the $U(1)_Y$
and $U(1)_X$ gauge bosons. After transforming to the canonical gauge kinetic terms, 
this mixing leads to an overall shift in the $U(1)_X$ charges 
as well as the coupled renormalization group equations for two $U(1)$ gauge couplings
and the $U(1)_X$ charge shift\cite{mixing}. 
However, when the extra gauge boson gets a heavy mass
for giving the see-saw scale for neutrino masses, 
the mixing effect is not relevant for the low energy physics while 
 the running of the gauge coupling for a light $U(1)$ gauge boson 
is not affected by the presence of the mixing term\cite{mixing}.  

Consequently, 
compared to eq.~(\ref{bmassless}), we obtain the relation between beta
functions as
\be
b_a=(0,-4,-6)+b^{++}_a+b^{-+}_a+b^{+-}_a+b^{--}_a.
\ee
The first term is only due to the difference between the beta functions
of ${\cal N}=1$ vector multiplets and ${\cal N}=2$ vector multiplets
for the SM gauge group. Apart from that, the sum of the ${\cal N}=2$ beta
functions for the volume dependent part
of the KK massive contributions, i.e. $\sum_{ij}b^{ij}_a$, contains 
only the KK massive modes for the bulk fields containing the zero modes.
Therefore, from the beta functions (\ref{b10}), (\ref{b16}), (\ref{b-+})
and (\ref{b--}),
one can find that the terms proportional to $\eta^R_1$ or $\eta^R_1\eta^R_2$,
i.e. the orbifold actions
associated with Pati-Salam and flipped $SU(5)$ gauge groups
generate the non-universal corrections to the gauge couplings.

From the obtained beta functions (\ref{bmassless}) 
and (\ref{b++})-(\ref{b--}),
eq.~(\ref{grun0}) becomes 
\bea
\frac{4\pi}{g^2_{{\rm eff}, a}(k^2)}&=&
\frac{4\pi}{g^2_u}
+\frac{1}{4\pi}{\tilde b}_a\ln\frac{M^2_*}{M^2_{B-L}}
+\frac{1}{4\pi}b'_a\ln\frac{M^2_{B-L}}{k^2} 
-\frac{1}{4\pi}\sum_{i,j=\pm}b^{ij}_a L_{ij}
+\frac{\Delta_a}{2\pi}.\label{grun}
\eea
Here ${\tilde b}_a=b_a-c_a+{\hat b}_a$ is the ${\cal N}=1$ beta function
above the $B-L$ breaking scale
and $b^{ij}_a$ are the beta functions for the KK massive modes.

\subsubsection{The differential running of the gauge couplings}

For a number of hyper multiplets with arbitrary parities,
we assume that both vector-like particles (getting
brane masses of order the GUT scale) and brane-localized particles
fill GUT multiplets, i.e. $c_a$ and ${\hat b}_a$ are universal. 
Then, we get the general formula for the differential running
of gauge couplings as
\bea
\frac{1}{g^2_3}-\frac{12}{7}\frac{1}{g^2_2}+\frac{5}{7}\frac{1}{g^2_1}
&=&\frac{1}{8\pi^2}\bigg({\tilde b}\ln\frac{M_*}{M_{B-L}} 
-\frac{1}{2}b^{-+}L_{-+} 
-\frac{1}{2}b^{--}L_{--}\bigg)
+\frac{{\tilde \Delta}}{8\pi^2}\label{gutrel}
\eea
where 
\bea
{\tilde b}
&=&\frac{9}{7}-\frac{9}{14}\sum_{10}\eta^{10}_1
-\frac{15}{14}\sum_{16}\eta^{16}_1+\frac{3}{7}\sum_{16}\eta^{16}_1\eta^{16}_2,
\\ 
b^{-+}
&=&-\frac{9}{7}-\frac{9}{14}\sum_{10}\eta^{10}_1
-\frac{15}{14}\sum_{16}\eta^{16}_1, \\
b^{--}
&=&\frac{12}{7}+\frac{3}{7}\sum_{16}\eta^{16}_1\eta^{16}_2.
\eea
Thus, we find a general relation between coefficients as
\be
{\tilde b}=\frac{6}{7}+b^{-+}+b^{--}.\label{relation}
\ee
Then, from eq.~(\ref{gutrel}) with the relation (\ref{relation}),
we find the deviation from the 4D SGUT prediction
of the QCD coupling at $M_Z$,
i.e.  $\Delta\alpha_s\equiv\alpha^{KK}_s-\alpha^{SGUT,0}_s$ as
\bea
\Delta \alpha_s(M_Z)
&\approx&-\frac{1}{2\pi}\alpha^2_s(M_Z)\bigg\{{\tilde b}
\ln\frac{M_*}{M_{B-L}}-({\tilde b}-\frac{6}{7})\ln(M_*\sqrt{V}) \nonumber \\
&&\quad-\frac{1}{2}b^{-+}\ln\Big[\frac{e^{-2}}{4}\Big|\vartheta_1(\frac{1}{2}|iu)
\Big|^4u\Big] \nonumber \\
&&\quad -\frac{1}{2}({\tilde b}-\frac{6}{7}-b^{-+})
\ln\Big[\frac{e^{-2}}{4}\Big|\vartheta_1(\frac{1}{2}-\frac{1}{2}iu|iu)e^{-\pi u/4}
\Big|^4u\Big]+{\tilde \Delta}\bigg\}.\label{dev}
\eea
The first term corresponds to the contribution due to the extra particles
above the $B-L$ scale. The second term is the volume dependent
correction due to the KK massive modes while the third part containing
the theta functions is the shape dependent correction.
The last term ${\tilde \Delta}$ is
the effect of the brane-localized gauge couplings. 

When $u\sim 1$,
the shape dependent term is subdominant compared
to the other logarithmic terms.
As can be shown explicitly in the specific models,
the last term can be also ignored by making a strong coupling
assumption at the cutoff scale.
Then, the first two logarithms become a dominant contribution.
For ${\tilde b}(\tilde b-\frac{6}{7})>0$, we can see that
the individual logarithm can be large, being compatible with
the gauge coupling unification due to a cancellation.

Now we consider the behavior of our result in the 5D limit 
with $u=R_6/R_5\gg 1$.
In this case, the bulk gauge group becomes the Pati-Salam and
there remain only two fixed points
with the Pati-Salam group and the SM gauge group enlarged
with a $U(1)$ factor, respectively. 
Some relevant discussion on this limit has been made in Ref.~\cite{hebecker},
concerning the power-like threshold corrections.
Since $|\vartheta_1(z|iu)|\sim 2e^{-\pi u/4}|\sin(\pi z)|$ for $u\gg 1$,
the shape dependent terms could give a significant effect on the gauge
coupling unification by the non-universal power-like threshold corrections.
Thus, in this 5D limit, eq.~(\ref{dev}) becomes
\bea
\Delta \alpha_s(M_Z)
&\approx&-\frac{1}{2\pi}\alpha^2_s(M_Z)\bigg\{{\tilde b}
\ln\frac{M_*}{M_{B-L}}-({\tilde b}-\frac{6}{7})\ln(M_*\sqrt{V}) \nonumber \\
&&\quad-\frac{1}{2}({\tilde b}-\frac{6}{7})\ln(4e^{-2}u) 
+\frac{\pi}{2}b^{-+} u
+{\tilde \Delta}\bigg\}.\label{dev2}
\eea
Therefore, we can interpret that 
the effective 5D gauge coupling($1/g^2_5=1/(g^2_4 R_6)$)
also gets a power-like threshold
correction proportional to $u/R_6\sim 1/R_5$ which is set
by the mass scale of heavy gauge bosons
belonging to $SO(10)/SU(4)\times SU(2)_L\times SU(2)_R$.

On the other hand, when we take a different 5D limit for $u\ll 1$,
the bulk gauge group 
becomes $SU(5)\times U(1)_X$ and
there remain only two fixed points
with $SU(5)\times U(1)_X$ and the SM gauge group enlarged
with a $U(1)$ factor, respectively. 
In this case, the appearing power threshold corrections are universal 
so there is no power threshold correction in eq.~(\ref{dev}).

\subsection{Gauge coupling unification in some $SO(10)$ orbifold GUT models}

Now we are in a position
to apply our general formula (\ref{dev})  to particular cases
for the unification of the SM gauge couplings.
To this purpose, we consider some known $SO(10)$ models of embedding
the MSSM into the extra dimensions.
In the minimal model(: model I)\cite{abc42}
that contains Higgs fields in the bulk
for breaking $U(1)_{B-L}$ and the SM gauge group\footnote{In order to cancel the bulk anomalies due to one $\bf 45$, we need to add
in the bulk two ${\bf 10}$'s.
So, it is necessary to have two Higgs doublets of the $\bf 10$'s in the bulk
unlike in 5D case\cite{kkl}.
Moreover, in order to break the $U(1)_{B-L}$, we need one ${\bf 16}$
in the bulk. However, for cancellation of localized and bulk anomalies,
one needs
one $\bf \overline {16}$ and two more ${\bf 10}$'s.}, there are
4 ${\bf 10}$'s with parities $(\eta_1,\eta_2)$
such as $H_1=(+,+)$,
$H_2=(+,-)$, $H_3=(-,+)$ and $H_4=(-,-)$,
and one pair of $\bf 16$ and $\bf \overline {16}$
with parities $\Phi=(-,+)$, $\Phi^c=(-,+)$.
Then, the resulting massless modes are two doublet Higgs fields $H^c_1$ and $H_2$ from $H_1$ and $H_2$,
and $G^c_3,G_4, (D^c,N^c),(D,N)$ from $H_3,H_4,\Phi$ and $\Phi^c$ in order.
Moreover, each family of quarks and leptons is introduced as a $\bf 16$
being localized at the fixed point without $SO(10)$ gauge symmetry.
After the $B-L$ breaking via the bulk $\bf 16$'s
with $\langle N\rangle=\langle N^c\rangle\neq 0$,
neutrino masses are generated at the fixed points by a usual
see-saw mechanism. Moreover,
$G^c_3,G_4, (D^c,N^c),(D,N)$ can acquire masses
of order the $B-L$ breaking scale by the brane
superpotential\cite{abc42,abcflavor}
$W=\lambda N D G^c_3+\lambda' N^c D^c G_4$
for $\langle N\rangle=\langle N^c\rangle\neq 0$.
In this case, since $\sum_{10}\eta^{10}_1=0$, $\sum_{16}\eta^{16}_1
=\sum_{16}\eta^{16}_1\eta^{16}_2=-2$,
we get the values ${\tilde b}=\frac{18}{7},b^{-+}=b^{--}=\frac{6}{7}$
in eq.~(\ref{dev}). Thus, in the 5D limit, because   
$b^{-+}$ is nonzero in eq.~(\ref{dev2}), there exists an effective
5D power-like threshold correction to the QCD coupling so the threshold
correction is sensitive to the shape modulus.

We consider another 6D $SO(10)$ GUT model
where the realistic flavor structure of the SM
was discussed(: model II)\cite{abcflavor}.
In this case,
on top of the minimal model, there are more hyper multiplets:
2 ${\bf 10}$'s
such as $H_5=(-,+)$ and $H_6=(-,-)$,
and one pair of $\bf 16$ and $\bf \overline {16}$ with $\phi=(+,+)$
and $\phi^c=(+,+)$.
Then, there are additional zero modes
$G^c_5,G_6,L,L^c$ from $H_5,H_6,\phi$ and $\phi^c$ in order.
They are assumed to get brane masses of order the GUT scale.
Thus, the running
of gauge couplings between the GUT scale and the $B-L$ breaking scale
is the same as in the minimal model.
In this case,
since $\sum_{10}\eta^{10}_1=-2$, $\sum_{16}\eta^{16}_1
=\sum_{16}\eta^{16}_1\eta^{16}_2=0$,
we get the values ${\tilde b}=\frac{18}{7},b^{-+}=0$ and
$b^{--}=\frac{12}{7}$ in eq.~(\ref{dev}).
Thus, in the 5D limit, because $b^{-+}=0$ in eq.~(\ref{dev2}),
there is no effective 5D power-like threshold correction to the QCD coupling.

\begin{figure}[t]

\begin{center}

\epsfig{file=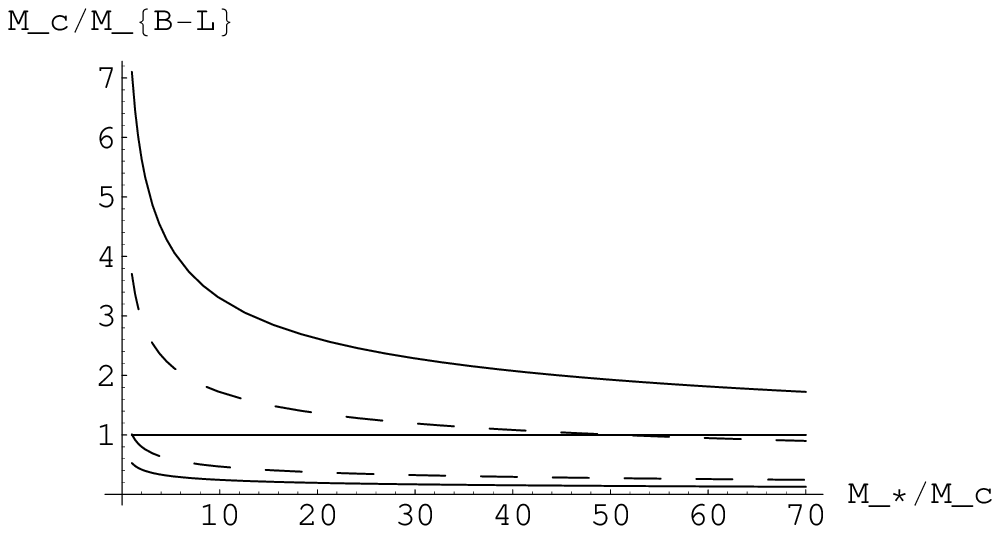,width=7cm,height=4cm}
\epsfig{file=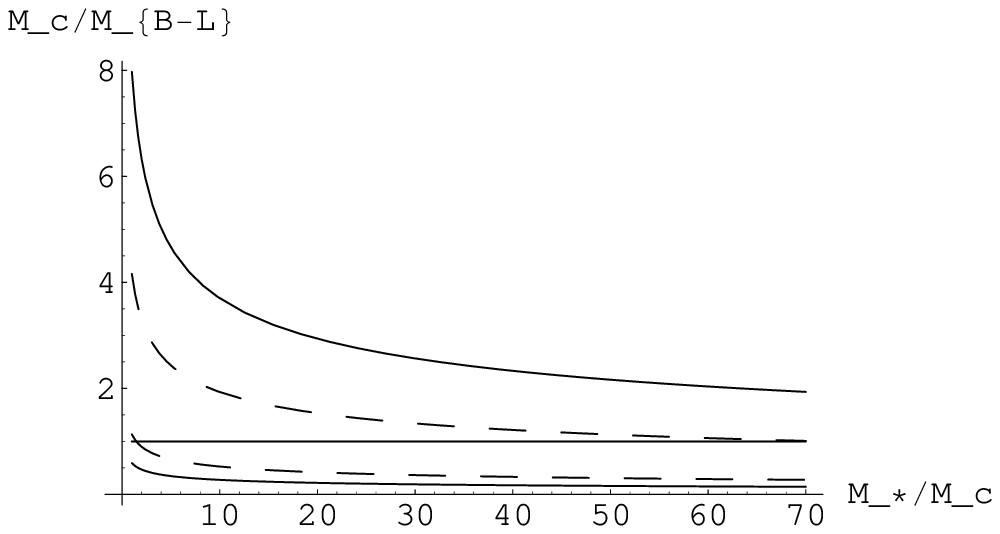,width=7cm,height=4cm}

\caption{The $1\sigma$ and $2\sigma$ band of $\Delta \alpha_s$: the model I
on the left and the model II on the right for $u=R_6/R_5\sim 1$.
The dashed lines and the continuous lines denote $1\sigma$ and $2\sigma$
bounds of the experimental data, respectively. Only in the region above the straight line at $M_c/M_{B-L}=1$, the $B-L$ breaking brane mass terms can be neglected. 
}
\end{center}
\label{alpha}

\end{figure}

From the data of the electroweak gauge couplings at the scale of the $Z$ mass,
one can compare the predicted value of the QCD coupling in a theory to a
measure one\cite{expstrong} $\alpha^{exp}_s=0.1176\pm 0.0020$.
In the 4D supersymmetric GUTs, the prediction without
threshold corrections for the QCD coupling
is $\alpha^{SGUT,0}_s=0.130\pm 0.004$. 
Thus, in this case, there is a discrepancy from the experimental data
as $\delta\alpha_s=\alpha^{SGUT,0}_s-\alpha^{exp}_s=0.0124\pm 0.0045$.

First we consider the case with isotropic compactification 
of the extra dimensions, $u\sim 1$.
In both models, since ${\tilde b}=\frac{18}{7}$, 
we can see from eq.~(\ref{dev}) that logarithmic contributions
of zero modes and those of KK massive modes appear with opposite signs
so that there is a possibility of having the large volume of extra
dimensions consistent with perturbativity and gauge coupling unification.
Ignoring the unknown brane-localized
gauge couplings and the $B-L$ breaking effect,
we depict in Fig.~1 the parameter space of $(M_c,M_{B-L})$ with
$u\sim 1$, being compatible with the experimental data.
If we take $M_*/M_c\sim 63/\sqrt{C}\sim 22$ for strong coupling 
assumption\footnote{We included the group theory factor $C=8$ 
for the $SO(10)$ bulk
gauge group in the naive dimensional analysis compared to \cite{naive}.}
at the 6D fundamental scale \cite{naive},
the correction due to the brane-localized gauge couplings
becomes ${\tilde \Delta}={\cal O}(1)$ so it is negligible to the KK threshold
corrections which is of order $\ln(M_*/M_c)\sim 3$.
In the model I(II), for $M_*/M_c\sim 22$, 
$M_{B-L}/M_c$ can be as small as $0.23(0.12)$ at the $2\sigma$ level. 
For $M_{B-L}/M_c\ll 1$, the KK massive modes of the color triplets are modified to
$m^2_{n_5,n_6}\approx(n_5/2R_5)^2+(n_6/2R_6)^2+c M^2_{B-L}$ where
$c$ is of order unity independent of the KK level for $R_5\neq R_6$\cite{dudas}.
In this case,
the $B-L$ breaking effect to the differential running is
estimated as $M^2_{B-L}/M^2_c$ in comparison to $\tilde \Delta$ in eq.~(\ref{dev}), 
so it is also suppressed compared to the KK threshold corrections.
Apart from the two models,
we can consider other possibilities of embedding the matter representations
into extra dimensions, like in the field-theory limit of a successful string
orbifold compactification\cite{buchmstring}
where there are two families at the fixed points and one family in the bulk.
In view of the general formula (\ref{dev}), however,
as far as an extra particle contributes to the running of the gauge couplings
above the $B-L$ breaking scale, $M_{B-L}$ tends to be close to $M_c$
for the success of the gauge coupling unification.

\begin{figure}

\begin{center}

\epsfig{file=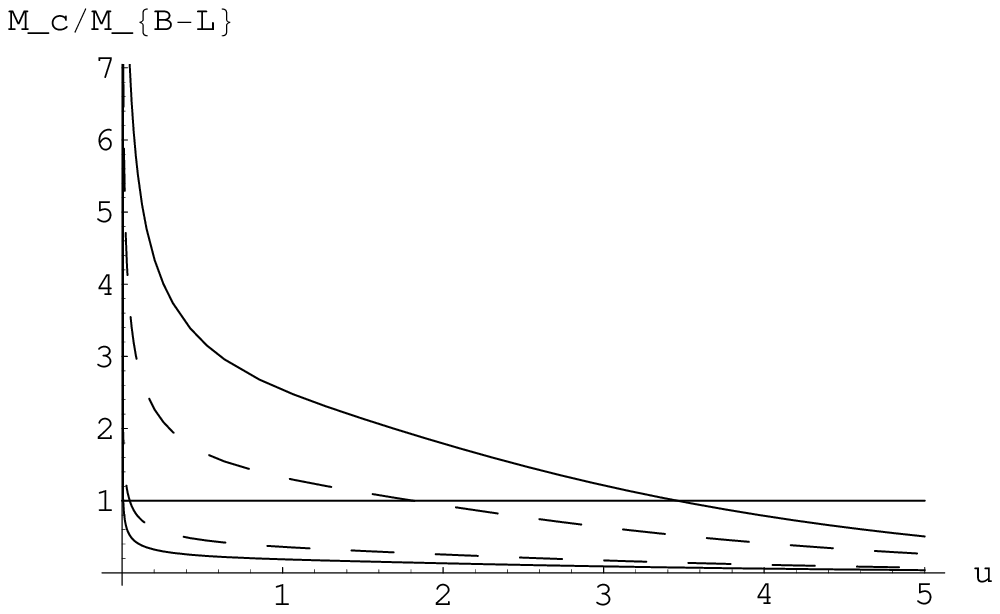,width=7cm,height=4cm}
\epsfig{file=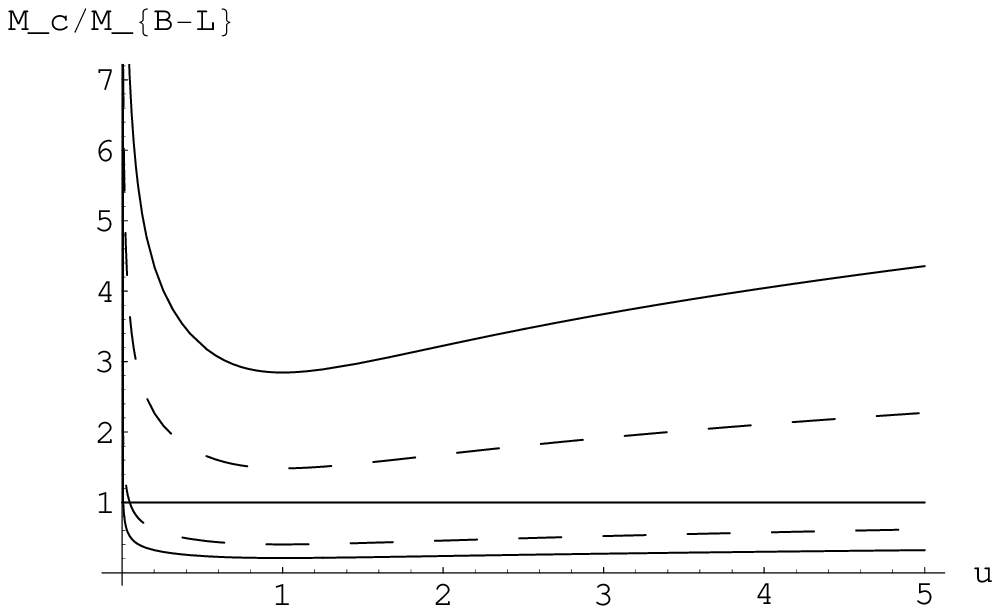,width=7cm,height=4cm}

\caption{The $1\sigma$ and $2\sigma$ band of $\Delta \alpha_s$: the model I
on the left and the model II on the right for $M_*/M_c\sim 22$.
The dashed lines and the continuous lines denote $1\sigma$ and $2\sigma$
bounds of the experimental data, respectively. Only in the region above the straight line at $M_c/M_{B-L}=1$, the $B-L$ breaking brane mass terms can be neglected. 
}
\end{center}
\label{beta}

\end{figure}

Next we consider the shape dependence of the loop corrections. 
As in the previous case, we make the strong coupling assumption 
and take the $B-L$ breaking scale to be below
the compactification scale. 
Then, we depict in Fig.~2 the parameter space of $(u,M_{B-L})$ with
$M_*/M_c\sim 22$, being compatible with the experimental data.
Therefore, we can see the clear difference between the two models, through the
dependence of the compatible $B-L$ breaking scale on the shape
modulus. As shown in the limit of anisotropic compactification of the extra 
dimensions in eq.~(\ref{dev2}), in the model I,  
the power-like threshold correction in the effective 5D theory gives a sizable  
contribution to the differential running of the gauge couplings, thus the $B-L$ breaking
scale is more sensitive to the shape modulus.

\section{Conclusion}
We have considered the one-loop effective action for gauge bosons
in six-dimensional orbifold GUT models with a number of hyper multiplets satisfying 
arbitrary local discrete twists.    
From the obtained effective action, we encountered the divergences which 
require the introduction of brane-localized gauge kinetic terms and a bulk higher derivative term.
Moreover, we derived the general expressions
for the running of the gauge couplings of zero-mode gauge bosons in the low 4D momentum limit.
Since the KK massive mode corrections depend on the parity actions,
in general they can give a non-universal contribution to the gauge coupling running. 

By taking a concrete example such as the $SO(10)$ orbifold GUTs,
we estimated the corresponding KK massive mode corrections to the QCD coupling 
at the scale of the $Z$ mass. The extra $U(1)$ or the $B-L$ symmetry 
is broken by the VEV of bulk singlets below the unification scale. 
Then, the extra color triplets, which accompany bulk singlets for a full $\bf 16$, 
also appear as zero modes so that they can lead to an additional logarithmic running of the gauge couplings
starting at the $B-L$ breaking scale. 

In the case with isotropic compactification of the extra dimensions, there is a partial
cancellation between two dominant logarithmic corrections; the volume dependent
part of the KK threshold corrections and the threshold corrections due to the extra color triplets.
In this case, we argued that 
the large volume of the extra dimensions can be compatible with the gauge coupling
unification and perturbativity. We also considered the case with anisotropic compactification
of the extra dimensions for which a 5D orbifold GUT limit can be discussed.
In this case, the shape dependent part of the KK threshold corrections corresponds to
non-universal power-like corrections in the compactification scales.
These power-like corrections are calculable because they are finite in the 6D sense.  
This situation is in contrast to the genuine 5D orbifold GUTs with non-simple groups 
where power-like corrections with the cutoff dependence is uncalculable. 
Further, we showed that 
for a fixed volome of the extra dimensions compatible with a strong coupling assumption, 
the $B-L$ breaking scale is sensitive to the change of the shape modulus, in order to maintain the success of the gauge coupling unification.

\vskip1cm

\section*{Acknowledgments}
The author would like to thank D.~Ghilencea and K.~Schmidt-Hoberg for early collaboration 
on the related topic. This work is supported in part by the DOE Contracts DOE-ER-40682-143
and DEAC02-6CH03000.

%%%%%%%%%%%%%%%%%%%%%%%%%%%%%%%%%%%%%%%%%%%%%%%%%%%%%%%%%%%%%%%%
\newpage

%%%%%%%%%%%%%%%%%%%%%%% Appendix %%%%%%%%%%%%%%%%%%%%%%%%%%%%%%%%

\def\theequation{A.\arabic{equation}}
\setcounter{equation}{0}
\vskip0.8cm
\noindent
{\Large \bf Appendix A: Propagators on GUT orbifolds}
\vskip0.4cm
\noindent

Suppose that the bulk gauge bosons satisfy the boundary conditions,
\bea
{\cal P}_i A_\mu(z){\cal P}^{-1}_i
\equiv P_i A_\mu(-z+z_i)P_i^{-1}&=&A_\mu(z+z_i), \\
{\cal P}_i A_m(z) {\cal P}^{-1}_i
\equiv -P_i A_m(-z+z_i)P_i^{-1}&=&A_m(z+z_i)
\eea
where $P^2_i=1(i=0,1,2,3)$ and $[P_i,P_j]=0$.
Then, we can write the gauge bosons ${\tilde A}_M$
on orbifolds in terms of gauge fields $A_M$ having $4\pi R_{5,6}$
periodicities along the extra dimensions as
\bea
{\tilde A}_M(z)=\prod_i\bigg[\frac{1}{2}(1+{\cal P}_i)\bigg]
A_M(z)\prod_j\bigg[\frac{1}{2}(1+{\cal P}_j^{-1})\bigg].
\eea
After taking into account the relation (\ref{p3}),
the field redefinition becomes in terms of component fields
in the group space
\bea
{\tilde A}^a_\mu(z)&=&\frac{1}{8}\bigg[\delta^a\,_b A^b_\mu(z)
+\sum_{i=0}^2(Q_i)^a\,_b
A^b_\mu(-z+2z_i)+\sum_{i<j\neq 3}(Q_i Q_j)^a\,_b
A^b_\mu(z+2z_i-2z_j) \nonumber \\
&&+(Q_0Q_1Q_2)^a\,_b
A^b_\mu(-z+2z_0-2z_1+2z_2)\bigg], \\
{\tilde A}^a_m(z)&=&\frac{1}{8}\bigg[\delta^a\,_b A^b_m(z)-
\sum_{i=0}^2(Q_i)^a\,_b
A^b_m(-z+2z_i)+\sum_{i<j\neq 3}(Q_i Q_j)^a\,_b
A^b_m(z+2z_i-2z_j) \nonumber \\
&&-(Q_0Q_1Q_2)^a\,_b
A^b_m(-z+2z_0-2z_1+2z_2)\bigg]
\eea
where $(Q_i)^a\,_b\equiv {\rm tr}(T^aP_iT_bP_i)$. 
Thus, the orbifold-compatible functional differentiations
for gauge fields are
\bea
(\delta^{{\tilde A}_{\mu(m)}}_{12})^a\,_b &=&
\frac{1}{8}\bigg[\delta^a\,_b \delta^2(z_1-z_2)
\pm \sum_{i=0}^2(Q_i)^a\,_b\delta^2(z_1+z_2-2z_i) \nonumber \\
&&+\sum_{i<j\neq 3}(Q_i Q_j)^a\,_b
\delta^2(z_1-z_2-2z_i+2z_j)\nonumber \\
&&\pm (Q_0Q_1Q_2)^a\,_b
\delta^2(z_1+z_2-2z_0+2z_1-2z_2)\bigg]\cdot \delta^4(x_1-x_2).
\eea
Consequently, the propagator of gauge fields in the Feynman gauge is given by
\bea
\langle {\tilde A}^a_M(z_1){\tilde A}_{b N}(z_2)\rangle
&=&g_{MN}(\delta^{{\tilde A}_M}_{13})^a\,_c G(z_3-z_4)
(\delta^{{\tilde A}_N}_{24})^c\,_b
\eea
where $G(z_3-z_4)$ is a bulk scalar propagator satisfying $4\pi R_{5,6}$
periodicities. Then, we obtain the propagator of gauge fields in 6D momentum
space as
\bea
\langle {\tilde A}^a_\mu(z_1){\tilde A}_{b \nu}(z_2)\rangle&\rightarrow&
g_{\mu\nu}\frac{1}{4}(1+Q_0Q_1\cos(2p_5\pi R_5))^a\,_c
(1+Q_0Q_2\cos(2p_6\pi R_6))^c\,_d \nonumber \\
&&\quad\times\frac{i}{2}\frac{(\delta_{{\vec p},{\vec p}'}
+Q_0\delta_{{\vec p},-{\vec p}'})^d\,_b}{p^2-{\vec p}^2}\equiv
g_{\mu\nu}({\tilde G}_{g,+}(p,{\vec p},{\vec p}'))^a\,_b, \\
\langle {\tilde A}^a_m(z_1){\tilde A}_{b n}(z_2)\rangle&\rightarrow&
g_{mn}\frac{1}{4}(1+Q_0Q_1\cos(2p_5\pi R_5))^a\,_c
(1+Q_0Q_2\cos(2p_6\pi R_6))^c\,_d \nonumber \\
&&\quad\times\frac{i}{2}\frac{(\delta_{{\vec p},{\vec p}'}
-Q_0\delta_{{\vec p},-{\vec p}'})^d\,_b}{p^2-{\vec p}^2}\equiv
g_{mn}({\tilde G}_{g,-}(p,{\vec p},{\vec p}'))^a\,_b
\eea
where ${\vec p}=(p_5,p_6)=(n_5/(2R_5),n_6/(2R_6))$ with $n_5,n_6$ integers.
We note that the Wilson line effects encoded into $Q_1$ and $Q_2$
are factorized as
two matrices in front of the propagator on orbifolds without Wilson lines.

Next let us consider a complex scalar field in the fundamental representation,
satisfying orbifold boundary
conditions,
\be
{\cal P}_i{\tilde\phi}(z)\equiv
\eta_i P_i{\tilde\phi}(-z+z_i)={\tilde\phi}(z+z_i)
\ee
with $\eta_i=+1$ or $-1$.
Similarly, we can write the complex scalar field $\tilde\phi$ on orbifolds
in terms of a complex scalar field satisfying $4\pi R_{5,6}$ periodicities as
\be
{\tilde\phi}(z)=\prod_{i}
\bigg[\frac{1}{2}(1+{\cal P}_i)\bigg]\phi(z).
\ee
Then, we can write the above equation in terms of component fields as
\bea
{\tilde\phi}^a(z)&=&\frac{1}{8}\bigg[\delta^a\,_b\phi(z)
+\sum_{i=0}^2\eta_i(P_i)^a\,_b\phi(-z+2z_i)
+\sum_{i<j\neq 3}\eta_i\eta_j(P_i P_j)^a\,_b
\phi^j(z+2z_i-2z_j) \nonumber \\
&&+\eta_0\eta_1\eta_2(P_0P_1P_2)^a\,_b\phi^b(-z+2z_0-2z_1+2z_2)\bigg].
\eea
Therefore, the orbifold-compatible functional differentiation for a complex
scalar field is
\bea
(\delta^{\tilde\phi}_{12})^a\,_b
&=&\frac{1}{8}\bigg[\delta^a\,_b\delta^2(z_1-z_2)
+\sum_{i=0}^2\eta_i(P_i)^a\,_b\delta^2(z_1+z_2-2z_i)
\nonumber \\
&&+\sum_{i<j\neq 3}\eta_i\eta_j(P_i P_j)^a\,_b
\delta^2(z_1-z_2-2z_i+2z_j) \nonumber \\
&&+\eta_0\eta_1\eta_2(P_0P_1P_2)^a\,_b\delta^2(z_1+z_2-2z_0+2z_1-2z_2)\bigg]
\cdot\delta^4(x_1-x_2).
\eea
Consequently, the propagator of a complex scalar is given by
\be
\langle{\tilde\phi}^a(z_1){\overline{\tilde\phi}}_b(z_2)\rangle=
(\delta^{\tilde\phi}_{13})^a\,_c G(z_3-z_4)(\delta^{\tilde\phi}_{24})^c\,_b
\ee
or in 6D momentum space,
\bea
\langle{\tilde\phi}^a(z_1){\tilde\phi}_b(z_2)\rangle
&\rightarrow& \frac{1}{4}(1+\eta_0\eta_1P_0P_1\cos(2p_5\pi R_5))^a\,_c
(1+\eta_0\eta_2P_0P_2\cos(2p_6\pi R_6))^c\,_d \nonumber \\
&&\quad\times \frac{i}{2}\frac{(\delta_{{\vec p},{\vec p}'}
+\eta_0P_0\delta_{{\vec p},-{\vec p}'})^d\,_b}{p^2-{\vec p}^2}
\equiv ({\tilde G}_{h,+}(p,{\vec p},{\vec p}'))^a\,_b.
\eea
For scalar fields of other representations, we only have to replace the
parity matrices with the ones for corresponding representations.

Finally let us consider a bulk left-handed
fermion in the fundamental representation,
satisfying the boundary conditions,
\be
{\cal P}_i{\tilde\psi}(z)\equiv
i\xi_i \gamma^5 P_i{\tilde\psi}(-z+z_i)={\tilde\psi}(z+z_i)
\ee
with $\xi_i=+1$ or $-1$.
Following the similar procedure, the propagator of a bulk fermion
is given in 6D momentum space as
\bea
\langle{\tilde\psi}^a(z_1){\overline{\tilde\psi}}_b(z_2)\rangle
&\rightarrow& \frac{1}{4}(1+\xi_0\xi_1 P_0P_1\cos(2p_5\pi R_5))^a\,_c
(1+\xi_0\xi_2P_0P_2\cos(2p_6\pi R_6))^c\,_d \nonumber \\
&&\quad\times \frac{i}{2}\bigg\{\frac{\delta^d\,_b\delta_{{\vec p},{\vec p}'}}
{p\hsp+\gamma_5 p_5+p_6}
-\xi_0(P_0)^d\,_b\frac{\delta_{{\vec p},-{\vec p}'}}{p\hsp+\gamma_5 p_5+p_6}
i\gamma_5\bigg\} \nonumber \\
&&\quad\equiv ({\tilde D}_\psi(p,{\vec p},{\vec p}'))^a\,_b.
\eea
On the other hand, 
for a bulk right-handed gaugino in the adjoint representation,
the propagator in 6D momentum space
takes the above form with $\xi_i P_i$ replaced by $Q_i$
and $p\hsp+\gamma_5 p_5+p_6\rightarrow p\hsp+\gamma_5 p_5-p_6$.

%%%%%%%%%%%%%

\def\theequation{B.\arabic{equation}}
\setcounter{equation}{0}
\vskip0.8cm
\noindent
{\Large \bf Appendix B: KK summations in 6D orbifolds}
\vskip0.4cm
\noindent

We consider the following KK summation
(with $c\geq 0$, $a_{1,2}>0$, $0\leq c_{1,2}< 1$):
\begin{eqnarray}\label{M1v}
\cJ_0[c; c_1,c_2]\!&\equiv & \Gamma[\epsilon/2] \sum_{n_1,n_2\in\bZ}\! 
 \Big[\pi  [c+a_1 (n_1+c_1)^2+a_2
    (n_2+c_2)^2]\Big]^{-\epsilon/2}
\nonumber\\[6pt]
&=&
\!\!\!\sum_{n_1,n_2\in\bZ}\! 
\!\!\int_0^\infty
\frac{dt}{t^{1-\epsilon/2}}\, \,e^{-\pi \,t\, [c+a_1 (n_1+c_1)^2+a_2
    (n_2+c_2)^2]}. 
\end{eqnarray}

If   $0\!\leq \!c/a_1\!<\! 1$,
with notations $\gamma(n_1)\equiv {\sqrt{z(n_1)}}/{\sqrt a_2}-i\, c_2$;
and $z(n_1)\equiv c \!+\! a_1 (n_1\!+\!c_1)^2$,
$u\equiv\sqrt{a_1/a_2}$,
$s_{\tilde n_1}\! \equiv\! 2\pi \tilde n_1 \sqrt{c/a_1}$,
$\gamma_E=0.577216...$, we obtain \cite{gls} (in the text $a_1=1/R_5^2$,
$a_2=1/R_6^2$)

\begin{eqnarray}\label{M1(0)}
&&\! \!\cJ_0[c; c_1,c_2] 
=  \! \frac{\pi c}{\sqrt{a_1 a_2}} \bigg[\frac{-2}{\epsilon}\!+\!
\ln\Big[4\pi \,a_1\,
e^{\gamma_E+\psi({c_1})+\psi(-{c_1})}\Big]\bigg] \!+ 2\pi\, u\,
\bigg[\frac{1}{6}+c_1^2-\big(c/a_1+c_1^2\big)^\frac{1}{2}\bigg]
\nonumber\\[6pt]
&&\!\!\!\!\!\!-\!\! \! \sum_{n_1\in\bZ}
\ln\Big\vert 1\!-\!e^{-2\pi \,\gamma (n_1)}\Big\vert^2
\!\!\! +\! \sqrt\pi\, u\!
\sum_{p\geq 1} \frac{\Gamma[p\!+\!1/2]}{(p\!+\!1)!}
\bigg[\frac{-c}{a_1}\bigg]^{p+1}
\!\!\!\Big(\zeta[2p\!+\!1,1\!+\!c_1]\!+\!\zeta[2p\!+\!1,1\!-\!c_1]
\!\Big)\qquad\,\,\,
\\[8pt]
\nonumber
\end{eqnarray}
while if  we have   $c/a_1>1$,  then

\begin{eqnarray}\label{M1(0)s}
\,\,\cJ_0[c; c_1,c_2] &=&
\frac{\pi c}{\sqrt{a_1 a_2}}\bigg[\! \frac{-2}{\epsilon}\!+\!
\ln\Big[
\!\pi \,c \,e^{\gamma_E-1}\Big]\bigg]
\!\! -\!\!\!\sum_{n_1\in\bZ}\!\!\ln\Big\vert
 1\!-\!e^{-2\pi \,\gamma (n_1)}\Big\vert^2\nonumber
\qquad\qquad\qquad\qquad\\[8pt]
&&\qquad\qquad\qquad+ \,
\frac{4\sqrt c}{\sqrt a_2}\sum_{\tilde n_1>0}
\frac{\cos(2\pi \tilde n_1\,c_1)}{\tilde n_1}\,
K_1(s_{\tilde n_1})\\[-4pt]\nonumber
\end{eqnarray}
The pole structure is the same for both cases; if $c/a_1>1$
and except the first square bracket,
 no power-like terms in $c$ are present (the last one being suppressed
due to $K_1$). Here $\zeta[z,a]$ is the Hurwitz  
Zeta function and  $\psi(x)=d/dx \,\ln \Gamma[x]$
and $K_1$ is the modified Bessel function. 

Finally, we quote here a limiting case for the behaviour of the
function $\cJ_0$
\begin{eqnarray}\label{limit-j0}
\cJ_0[c\ll 1; 0,0]
 &=&  \frac{\pi c}{\sqrt{a_1 a_2}} \bigg[\frac{-2}{\epsilon}
+\ln \Big[4\pi e^{-\gamma_E}
a_1\big|\eta(i\sqrt{a_1/a_2})\big|^4\Big]\bigg] \nonumber \\
&&-\ln\Big[ 4\pi^2 \, \vert\eta(i \, \sqrt{a_1/a_2}) \vert^4
\,\,a_2^{-1}\Big]-\ln c \nonumber \\
\cJ_0[c\ll 1;0,1/2]
&=& \frac{\pi c}{\sqrt{a_1 a_2}}\bigg[\frac{-2}{\epsilon}
+\ln(4\pi e^{-\gamma_E}a_1)-\pi u+2\sum_{n\geq 1}(-1)^n\ln|1-e^{-2\pi un}|^2
\bigg] \nonumber \\
&&-\ln\bigg|\frac{\vartheta_1(1/2|iu)}{\eta(iu)}\bigg|^2 \nonumber \\
\cJ_0[c\ll 1; 1/2,0]&=&\frac{\pi c}{\sqrt{a_1 a_2}}
\bigg[\frac{-2}{\epsilon}+\ln(\pi e^{-\gamma_E}a_1)
-2\sum_{n\geq 1} 
\ln\bigg|\frac{1+e^{-\pi un}}{1-e^{-\pi un}}\bigg|^2\bigg] \nonumber \\
&&-\ln\bigg\vert\frac{\vartheta_1(-i u/2\vert i u)}{\eta(i u) }e^{-\pi u/4}\bigg\vert^2
\nonumber\\
\cJ_0[c\ll 1; 1/2,1/2] 
& =&  \frac{\pi c}{\sqrt{a_1 a_2}} \bigg[\frac{-2}{\epsilon}
+\ln(\pi e^{-\gamma_E}a_1)-2\sum_{n\geq 1}(-1)^n
\ln\bigg|\frac{1+e^{-\pi un}}{1-e^{-\pi un}}\bigg|^2\bigg] 
\nonumber \\
&&-\ln\bigg\vert\frac{\vartheta_1(1/2-i u/2\vert i u)}{\eta(iu)} e^{-\pi u/4}\bigg\vert^2.
\end{eqnarray}

\def\theequation{C.\arabic{equation}}
\setcounter{equation}{0}
\vskip0.8cm
\noindent
{\Large \bf Appendix C: Definition of special functions}
\vskip0.4cm
\noindent

The Hurwitz Zeta function $\zeta[z,a]$ is defined as 
\be
\zeta[z,a]=\sum_{n\geq 0} (n+a)^{-z}
\ee
with $\rm{Re}\,z>1$ and $a\not=0,-1,-2,\cdots$.

The modified Bessel function $K_n$ is defined through 
\begin{equation}\label{bessel1}
\int_{0}^{\infty} \! dx\, x^{\nu-1} e^{- b x^p- a
x^{-p}}=\frac{2}{p}\, \bigg[\frac{a}{b}
\bigg]^{\frac{\nu}{2 p}} K_{\frac{\nu}{p}}(2 \sqrt{a \, b}),\quad Re
(b),\, Re (a)>0
\end{equation}
with
\begin{eqnarray}
K_1[x]=e^{-x}\sqrt{\frac{\pi}{2
    x}}\left[1+\frac{3}{8 x}-\frac{15}{128 x^2}
+\cO(1/x^3)\right]
\\
\nonumber
\end{eqnarray}
which is  strongly suppressed at large argument.

In the text, we used the Dedekind Eta function 
\begin{eqnarray}
\eta(\tau) & \equiv & e^{\pi i \tau/12} \prod_{n\geq 1} (1- e^{2 i
\pi\tau\, n}), \nonumber \\
\eta(-1/\tau)&=&\sqrt{-i\tau}\,\eta(\tau), \quad  
\eta(\tau+1)=e^{i\pi/12}\eta(\tau),
\end{eqnarray}
and the Jacobi Theta function $\vartheta_1$
\begin{eqnarray}
\vartheta_1(z\vert\tau)&\equiv & 2 q^{1/8}\sin (\pi z) \prod_{n\geq 1}
(1- q^n) (1-q^n e^{2 i \pi z}) (1- q^n e^{-2 i \pi z}), \qquad
q\equiv e^{2 i \pi \tau} \nonumber \\
&=&-i\sum_{n\in {\bf Z}}(-1)^n e^{i\pi \tau(n+1/2)^2} e^{(2n+1)i\pi z}
\end{eqnarray}
which has the properties
\bea
\vartheta_1(z|\tau+1)&=&e^{i\pi/4}\vartheta_1(z|\tau), 
\nonumber \\
\vartheta_1(z+1|\tau)&=&-\vartheta_1(z|\tau), \nonumber\\
\vartheta_1(z+\tau|\tau)&=&-e^{-i\pi\tau-2i\pi z}\vartheta_1(z|\tau), 
\nonumber\\
\vartheta_1(-z/\tau|-1/\tau)&=&e^{i\pi/4}\tau^{1/2}e^{i\pi z^2/\tau}
\vartheta_1(z|\tau).
\eea

\def\theequation{D.\arabic{equation}}
\setcounter{equation}{0}
\vskip0.8cm
\noindent
{\Large \bf Appendix D: Some group theory for $SO(10)$ GUT}
\vskip0.4cm
\noindent

\begin{itemize}
\item Relations between fundamental and other representations:

For $SU(N)$ and $SO(2N)$ gauge groups considered in the paper,
we have \cite{erler}
\bea
{\rm tr}_{\rm Adj}F^2_{SU(N)}&=&2N \,{\rm tr}_N F^2_{SU(N)}, \\
{\rm tr}_{a^{ij}}F^2_{SU(N)}&=&(N-2)\,{\rm tr}_N F^2_{SU(N)}, \\
{\rm tr}_{a^{ijk}}F^2_{SU(N)}&=&\frac{1}{2}(N-2)(N-3)\,{\rm tr}_N F^2_{SU(N)}, 
\\
{\rm tr}_{\rm Adj}F^2_{SO(2N)}&=&2(N-1)\,{\rm tr}_{2N} F^2_{SO(2N)}, \\
{\rm tr}_{2^{N-1}}F^2_{SO(2N)}&=&2^{N-4}\,{\rm tr}_{2N} F^2_{SO(2N)}
\eea
where the subindex of the trace implies the representation of the group,
for instance, $a^{ij}(a^{ijk})$ is the second(third) 
rank totally antisymmetric tensor representation of $SU(N)$. 
In the text, we take the normalization, 
${\rm tr}_N (T^a T^b)=\frac{1}{2}\delta^{ab}$ for $SU(N)$
and ${\rm tr}_{2N} (T^a T^b)=\delta^{ab}$ for $SO(2N)$.

\item Computation of traces: \\
By standard representation theory, we can do the decomposition of the quadratic
Casimir for an adjoint representation of $SO(10)$:
under the Pati-Salam,
\bea
{\rm tr}_{\rm Adj}F^2_{SO(10)}
&=&{\rm tr}_{\rm Adj}F^2_{SU(4)}+4{\rm tr}_6 F^2_{SU(4)}
+12{\rm tr}_2 F^2_{SU(2)_L}+12 {\rm tr}_2 F^2_{SU(2)_R} \nonumber \\
&&+{\rm tr}_{\rm Adj}F^2_{SU(2)_L}+{\rm tr}_{\rm Adj}F^2_{SU(2)_R}
\eea
and under the Georgi-Glashow,
\bea
{\rm tr}_{\rm Adj}F^2_{SO(10)}
&=&{\rm tr}_{\rm Adj}F^2_{SU(5)}+{\rm tr}_{10}F^2_{SU(5)}
+{\rm tr}_{\overline {10}}F^2_{SU(5)}+8F^2_{U(1)_X}.
\eea
Using the definition
${\rm tr}_{\rm Adj}(Q_iF^2_{SO(10)})=f_{acd}f_{bef}\eta^{cd}Q^{df}_iF^a_{SO(10)}
F^b_{SO(10)}$
and the fact that $Q_i$ is equal to $+1(-1)$ for $Z_2$-even(odd) modes of gauge fields,
we get
\bea
{\rm tr}_{\rm Adj}(Q_1F^2_{SO(10)})
&=&{\rm tr}_{\rm Adj}F^2_{SU(4)}-4{\rm tr}_6 F^2_{SU(4)}
-12{\rm tr}_2 F^2_{SU(2)_L}-12 {\rm tr}_2 F^2_{SU(2)_R} \nonumber \\
&&+{\rm tr}_{\rm Adj}F^2_{SU(2)_L}+{\rm tr}_{\rm Adj}F^2_{SU(2)_R} \nonumber
\\
&=&-8{\rm tr}_2 F^2_{SU(2)_L}-8{\rm tr}_2 F^2_{SU(2)_R}
\eea
and
\bea
{\rm tr}_{\rm Adj}(Q_2F^2_{SO(10)})
&=&{\rm tr}_{\rm Adj}F^2_{SU(5)}-{\rm tr}_{10}F^2_{SU(5)}
-{\rm tr}_{\overline {10}}F^2_{SU(5)}-8F^2_{U(1)_X} \nonumber \\
&=&4{\rm tr}_5F^2_{SU(5)}-8F^2_{U(1)_X}
\eea
and likewise
\be
{\rm tr}_{\rm Adj}(Q_3F^2_{SO(10)})=4{\rm tr}_5F^2_{SU(5)'}
-8F^2_{U(1)_{X'}}.
\ee

Let us consider a similar decomposition of the index of the other
representation of $SO(10)$.
First, the index of a fundamental representation of $SO(10)$ is decomposed
as
\bea
{\rm tr}_{10}F^2_{SO(10)}&=&{\rm tr}_6 F^2_{SU(4)}+2{\rm tr}_2 F^2_{SU(2)_L}
+2{\rm tr}_2 F^2_{SU(2)_R} \nonumber \\
&=&{\rm tr}_5 F^2_{SU(5)}+\frac{1}{2}F^2_{U(1)_X}
+{\rm tr}_{\overline 5} F^2_{SU(5)}
+\frac{1}{2}F^2_{U(1)_X}.
\eea
Then, with the parity matrices in the traces, we get
\bea
{\rm tr}_{10}(P_1F^2_{SO(10)})&=&-{\rm tr}_6 F^2_{SU(4)}
+2{\rm tr}_2 F^2_{SU(2)_L}+2{\rm tr}_2 F^2_{SU(2)_R} \nonumber \\
&=&-2{\rm tr}_4F^2_{SU(4)}+2{\rm tr}_2 F^2_{SU(2)_L}+2{\rm tr}_2 F^2_{SU(2)_R}
\eea
and
\bea
{\rm tr}_{10}(P_2 F^2_{SO(10)})&=&-{\rm tr}_5 F^2_{SU(5)}
-\frac{1}{2}F^2_{U(1)_X}
+{\rm tr}_{\overline 5} F^2_{SU(5)}
+\frac{1}{2}F^2_{U(1)_X}=0
\eea
and similarly ${\rm tr}_{10}(P_3F^2_{SO(10)})=0$.
Next, we also do the decomposition of the index
of a $\bf 16$ spinor representation of $SO(10)$ as
\bea
{\rm tr}_{16}F^2_{SO(10)}&=&2{\rm tr}_4 F^2_{SU(4)}
+2{\rm tr}_{\overline 4} F^2_{SU(4)}+4 {\rm tr}_2 F^2_{SU(2)_L}
+4{\rm tr}_2 F^2_{SU(2)_R} \nonumber \\
&=&{\rm tr}_{10} F^2_{SU(5)}+{\rm tr}_{\overline 5} F^2_{SU(5)}
+\frac{1}{40}(10+45+25)F^2_{U(1)_X}.
\eea
Then, we get the necessary traces for a $\bf 16$ as
\bea
{\rm tr}_{16}(P_1F^2_{SO(10)})&=&2{\rm tr}_4 F^2_{SU(4)}
-2{\rm tr}_{\overline 4} F^2_{SU(4)}+4 {\rm tr}_2 F^2_{SU(2)_L}
-4{\rm tr}_2 F^2_{SU(2)_R} \nonumber \\
&=&4{\rm tr}_2 F^2_{SU(2)_L}-4{\rm tr}_2 F^2_{SU(2)_R}
\eea
and
\bea
{\rm tr}_{16}(P_2F^2_{SO(10)})&=&-{\rm tr}_{10} F^2_{SU(5)}
+{\rm tr}_{\overline 5} F^2_{SU(5)}+\frac{1}{40}(-10+45+25)F^2_{U(1)_X}
\nonumber \\
&=&-2{\rm tr}_5 F^2_{SU(5)}+\frac{3}{2}F^2_{U(1)_X}
\eea
and likewise
\be
{\rm tr}_{16}(P_3F^2_{SO(10)})
=-2{\rm tr}_5 F^2_{SU(5)'}+\frac{3}{2}F^2_{U(1)_{X'}}.
\ee

\item Relations between $U(1)$ generators: \\ 
There are three maximal subgroups of $SO(10)$, 
Georgi-Glashow ($SU(5)\times U(1)_X$) and 
Pati-Salam ($SU(4)\times SU(2)_L\times SU(2)_R$)
and flipped $SU(5)$ ($SU(5)'\times U(1)_{X'}$). 
For a fundamental representation of $SO(10)$, the $U(1)$ generators are given
by
\bea
&&Y={\rm diag}(\frac{1}{3},\frac{1}{3},\frac{1}{3},-\frac{1}{2},-\frac{1}{2})\times \sigma^2, \ \
X={\rm diag}(2,2,2,2,2) \times\sigma^2, \nonumber \\
&&B-L={\rm diag}(\frac{2}{3},\frac{2}{3},\frac{2}{3},0,0)\times \sigma^2, \ \
T_{3R}={\rm diag}(0,0,0,-\frac{1}{2},-\frac{1}{2})\times \sigma^2.
\eea
Thus, we obtain the relation between $U(1)$ generators appearing
in the different subgroups,
\bea
Y=T_{3R}+\frac{1}{2}(B-L), \ \
X=-4T_{3R}+3(B-L). \label{g1}
\eea
Moreover, by comparing the flipped $SU(5)$ to the
Georgi-Glashow as $N\leftrightarrow e^c_L$  
and $u^c_L\leftrightarrow d^c_L$ in $\bf 16$ and 
$h_1\leftrightarrow h_2$ in $\bf 10$, 
we get another relation
\bea
Y=\frac{1}{5}(-Y'+X'), \ \  
X=\frac{1}{5}(24Y'+X'). \label{g2}
\eea
Using the above relations, we can also find the relations between the $U(1)$
gauge bosons. To this, let us consider the bulk kinetic terms for $U(1)$
gauge bosons $A_Y,A_X$ and a charged field $\phi$:
\be
{\cal L}_{\rm bulk}\supset -\frac{1}{4g^2_1}F^2_Y-\frac{1}{4g^2_X}
F^2_X
+\Big|\Big[\partial-i(\sqrt{\frac{3}{5}}YA_Y
+\frac{1}{\sqrt{40}}XA_X)\Big]\phi\Big|^2
\ee
where $g_1=g_X$ at tree level. 
By writing  
\bea
\sqrt{\frac{3}{5}}YA_Y+\frac{1}{\sqrt{40}}XA_X &=&
\sqrt{\frac{3}{5}}Y'A_{Y'}+\frac{1}{\sqrt{40}}X'A_{X'} \nonumber \\
&=&T_{3R}A_R+\sqrt{\frac{3}{8}}(B-L) A_{B-L},
\eea
and using eqs.~(\ref{g1}) and (\ref{g2}),
we obtain the relations between the $U(1)$ gauge bosons as
\bea
A_{Y'}&=&\frac{1}{5}(-A_Y+2\sqrt{6}A_X), \ \
A_{X'}=\frac{1}{5}(2\sqrt{6}A_Y+A_X), \nonumber \\  
A_R&=&\sqrt{\frac{3}{5}}(A_Y-\sqrt{\frac{2}{3}}A_X), \ \
A_{B-L}=\sqrt{\frac{3}{5}}(\sqrt{\frac{2}{3}}A_Y+A_X). \label{u1grel}
\eea

\end{itemize}

%%%%%%%%%%%%%%%%%%%%%%%%%%%%%

\end{document}